\documentclass{iopart}

\usepackage{iopams} 
\usepackage{amstext}

\usepackage{graphicx}%
\usepackage{dcolumn}%
\usepackage{mathrsfs}
\usepackage{bm}
\usepackage{color,xspace}
\usepackage[small,compact,raggedright]{titlesec}
\usepackage{soul,hyperref}
\usepackage{iopams}
\usepackage{setstack}
\usepackage{graphicx}
\usepackage{xspace}
\usepackage{tikz}
\usepackage{alltt}

\newcommand{\eqref}[1]{{(\ref{#1})}}
\newcommand{\da}{{\Box_g}}

\newcommand{\raro}[1]{\mathcal #1}

\newcommand{\ket}[1]{\left| {#1} \right\rangle}
\newcommand{\bra}[1]{\left\langle {#1} \right|}

\newcommand{\proj}[2]{\left| {#1} \right\rangle\!\left\langle {#2} \right|}

\newcommand{\ii}{\mathrm{i}}
\newcommand{\ee}{\mathrm{e}}
\newcommand{\textmod}[1]{\mathrm{#1}}
\def\b{\begin{equation}}
\def\e{\end{equation}}

\newcommand{\lplanck}{{\ell_{\mathrm{p}}}}

\def\reals{\mathbb{R}}

\renewcommand{\vec}[1]{\bm{\mathrm{#1}}}
\renewcommand{\mat}[1]{\bm{\mathrm{#1}}}
\newcommand{\op}[1]{\hat{#1}}
\newcommand{\opvec}[1]{\op{\vec{#1}}}

\newcommand{\tp}{\mathrm{T}}
\newcommand{\herm}{\mathrm{H}}
\newcommand{\coloneqq}{\mathrel{:=}}

\makeatletter
\@ifundefined{textcolor}{}
{%
 \definecolor{BLACK}{gray}{0}
 \definecolor{WHITE}{gray}{1}
 \definecolor{RED}{rgb}{1,0,0}
 \definecolor{GREEN}{rgb}{0,.4,0}
 \definecolor{BLUE}{rgb}{0,0,1}
 \definecolor{CYAN}{cmyk}{1,0,0,0}
 \definecolor{MAGENTA}{cmyk}{0,1,0,0}
 \definecolor{YELLOW}{cmyk}{.2,.4,1,0}
 }
\makeatother

\begin{document}

\title{Entanglement in curved spacetimes and cosmology}

\author{Eduardo Mart\'in-Mart\'inez}

\address{Institute for Quantum Computing, University of Waterloo, Waterloo, Ontario, N2L 3G1, Canada}
\address{Department of Applied Mathematics, University of Waterloo, Waterloo, Ontario, N2L 3G1, Canada}
\address{Perimeter Institute for Theoretical Physics, Waterloo, Ontario, N2L 6B9, Canada}

\author{Nicolas C. Menicucci}

\address{School of Physics, The University of Sydney, Sydney, New South Wales, 2006, Australia}

\begin{abstract}
We review recent results regarding entanglement in quantum fields in cosmological spacetimes and related phenomena in flat spacetime such as the Unruh effect. We being with a summary of important results about field entanglement and the mathematics of Bogoliubov transformations that is very often used to describe it. We then discuss the Unruh-DeWitt detector model, which is a useful model of a generic local particle detector. This detector model has been successfully used as a tool to obtain many important results. In this context we discuss two specific types of these detectors: a qubit and a harmonic oscillator. The latter has recently been shown to have important applications when one wants to probe nonperturbative physics of detectors interacting with quantum fields. We then detail several recent advances in the study and application of these ideas, including echoes of the early universe, entanglement harvesting, and a nascent proposal for quantum seismology.
\end{abstract}

\maketitle

\section{Introduction}

It is a rather alarming fact that nearly all of the energy of the universe is so-called \emph{dark energy}---i.e.,~energy of an unknown nature that is nevertheless driving an unabated and accelerating expansion of space and everything in it~\cite{Riess1998,Perlmutter1999}. Furthermore, even considering just the matter content of the universe, the vast majority of it is unexplained, as well. Measurements of the deuterium-hydrogen ratio in interstellar absorption, together with some theoretical models of cosmological nucleosynthesis, have provided evidence that unspecified, non-baryonic matter accounts for an astounding 80\% of the matter content of the universe~\cite{Reid2010}. The nature of this \emph{cold dark matter} cannot be explained exclusively with standard-model neutrinos~\cite{neutrinoref} nor do recent results in experimental particle physics from the Large Hadron Collider~\cite{LHC1,LHC2,LHC3,LHC4,LHC5} seem to shed any light on the matter.

With the vast majority of the universe being of unknown origin and driving its own  accelerating expansion, we are further stymied by the fact that we do not fully understand the nature of the quantum field responsible for the inflation period in the very early universe~\cite{Peiris2003}. Nevertheless, the BICEP2 experiment has recently  claimed discovery of  the signature of primordial gravitational waves imprinted onto the polarization of the cosmic microwave background (CMB)~\cite{Collaboration:2014wq}. If confirmed in further experiments, this would constitute the first experimental discovery of primordial gravitational waves corresponding  to quantum fluctuations of the metric itself. This would also constitute the second indirect observation of gravitational radiation. Indeed gravitational waves have not been observed as of the publication of this review, and we only have indirect proof of their existence through their effects on the orbital dynamics of binary pulsars~\cite{GravWavesBin}. Furthermore, we still have a long way to go before relativity can be successfully married to quantum field theory in a full theory of quantum gravity---i.e.,~one that necessarily includes the backreaction of quantised matter on spacetime.

In the limit where this backreaction is negligible, quantum field theory in curved spacetime is the most complete theory so far~\cite{Birrell1982,Wald2}. Cosmological spacetimes provide a useful class of all curved backgrounds on which to study the behaviour of quantum fields since they are important throughout the entire life cycle of our universe, from inflation to recombination to dark energy-fueled expansion. Furthermore, even flat (Minkowski) spacetime provides a useful testbed for quantum fields in the low-curvature limit, which can still reveal interesting phenomena such as the Fulling-Davies-Unruh effect~\cite{Fulling1973,Davies1975,Unruh1976,Crispino2008} (often called just the Unruh effect) for accelerating observers. This effect, which results in accelerating observers detecting thermal radiation in the Minkowski vacuum~\cite{Unruh1976}, is mathematically related to an analogous effect in exponentially expanding (de~Sitter) spacetime: the Gibbons-Hawking effect~\cite{Gibbons1977}, which predicts a similar detector response due to the transient effects of an expanding universe.

In this article, we explore these effects from the perspective of quantum entanglement and relativistic quantum optics. We start with a brief review of the basic mathematical formalism in QFT in curved spacetimes. Then we review recent results that discuss the entanglement naturally occurring in a quantum field and produced during the expansion of spacetime. We also discuss  closely  related phenomena in the context of the Unruh effect for accelerating observers. Special attention is payed to the role of particle detectors and their ability to harvest entanglement from the underlying quantum field both in flat and cosmological spacetime backgrounds. We choose an Unruh-DeWitt detector model because of its ubiquity in the literature and because recent results show that using a harmonic oscillator  (in place of the more usual qubit)  as the local quantum system allows one to do nonperturbative analysis of detector behavior when interacting with free fields. This powerful technique opens the door to new results and possible practical applications, including a speculative proposal for quantum seismology using a technique known as entanglement farming. Furthermore, we discuss recent results suggesting how detectors today could, in principle, `hear the echo' of quantum-gravity features of the Big Bang and the early Universe.

\section{Bogoliubov transformations: Useful mathematical tools for quantum field theory in curved spacetimes}

\subsection{Non-inertial observers of quantum fields in flat spacetime}\label{Rindbogo}

In quantum field theory, different observers define different natural quantization schemes. In flat spacetime, all the quantization schemes corresponding to inertial observers are equivalent (they would all agree on what the field vacuum is). This is not true in general. Indeed, different geodesic observers in general spacetimes do not necessarily agree on what the vacuum of a quantum field theory is.

Even for the simple case of a scalar quantum field on flat spacetime, there are non-trivial differences between observers in different kinematic states. For instance, it is well-known that accelerated observers would not agree with inertial observers in their definition of the vacuum. To illustrate this and as a convenient introduction, we will briefly carry out a flat spacetime scalar field quantization  for two different class of observers of a quantum field, inertial and constantly accelerated.

To deal with uniformly accelerated observers, we introduce Rindler coordinates $(\tau,\xi)$~\cite{gravitation}, which are the proper coordinates of an observer moving at constant acceleration $a$. The correspondence between the proper coordinates of a stationary observer (chosen w.l.g. at rest) $(t,x)$ and the proper coordinates of an accelerated observer $(\tau,\xi)$ is 
\begin{equation}\label{change}
ct=\xi \sinh\left(\frac{a\tau}{c}\right),\qquad x=\xi\cosh\left(\frac{a\tau}{c}\right),
\end{equation}
where, we have made $c$ explicit\footnote{Note that we are not using  conformal Rindler coordinates $t=c\,a^{-1}e^{a\xi/c^2}\sinh\left(\frac{a\tau}{c}\right)$ and $x=c^{2}a^{-1}e^{a\xi/c^2}\cosh\left(\frac{a\tau}{c}\right)$ (quite common in the literature) but the proper coordinates of an accelerated observer of acceleration $a$ such that the proper lengths and times measured in these units are physical distances and time lapses as measured by the accelerated observer.}.  
\begin{figure}[ht]
\begin{center}
\includegraphics[width=.50\textwidth]{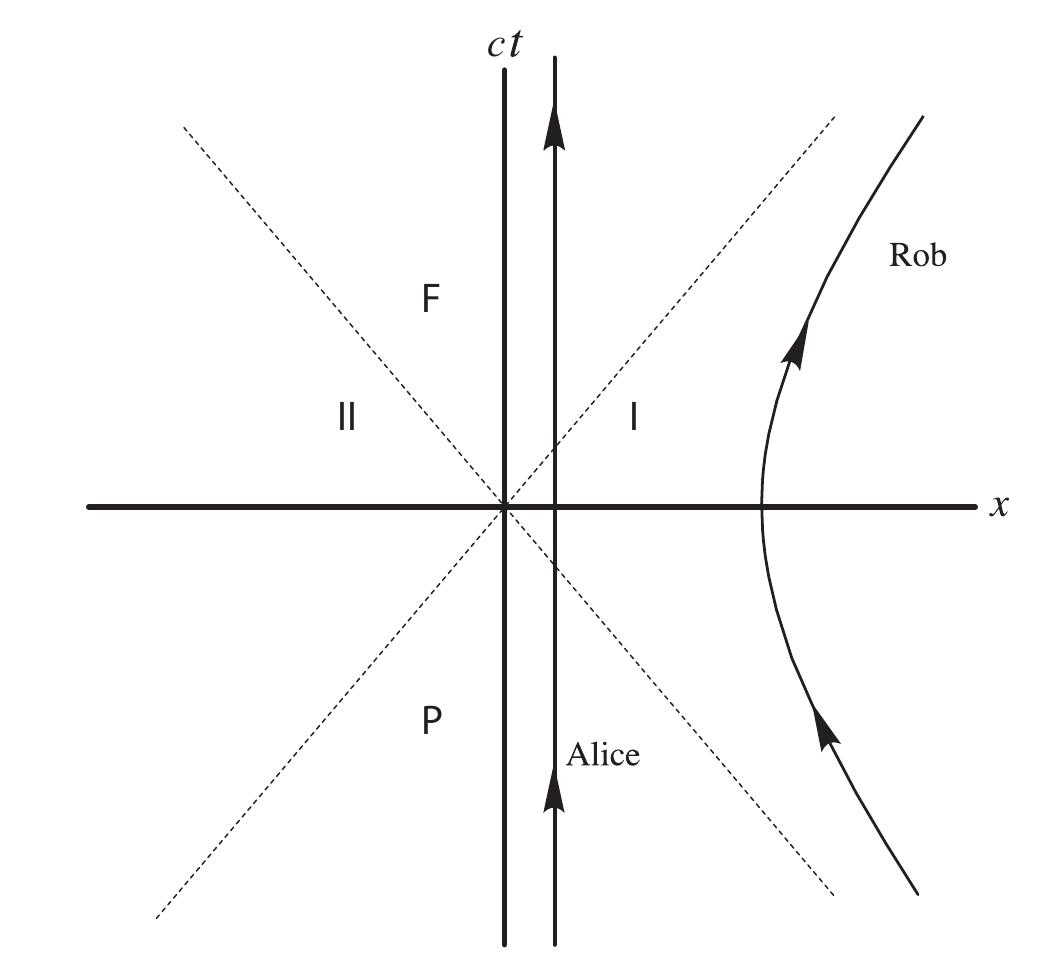}
\end{center}
\caption{Flat spacetime. Trajectories of an inertial (Alice) and constantly accelerated (Rob) observers}
\label{rin2}
\end{figure}

Directly from \eqref{change} we see that trajectories of observers with constant acceleration correspond to $\xi=\text{const.}$. These trajectories  are hyperbolae whose asymptote is the light cone (as the observer accelerates his velocity tends to the speed of light). By choosing a particular value of the parameter $a$ in \eqref{change} we are specifying the particular observer for which these coordinates correspond to his proper frame. To find this observer's Rindler position $\xi$ we can use that all Rindler observers are instantaneously at rest at time $t=0$ in the inertial frame, and at this time a Rindler observer with proper acceleration $a$ ---and therefore, proper coordinates $(\xi,\tau)$--- will be at Minkowskian position $x=c^2/a$. On the other hand, at this point $t=0\Rightarrow\xi=x$ instantaneously so, consequently, the constant Rindler position for this trajectory is $\xi= c^2/a$.

One can easily see that for eternally  accelerated observers, an acceleration horizon appears: Any accelerated observer would be restricted to either region I or II of  spacetime as per Fig. \ref{rin2}.   

Indeed, a quick inspection reveals that the Rindler coodinates defined in \eqref{change} do not cover the whole Minkowski spacetime, but only the right wedge (Region I in Figure \ref{rin2}). In fact to map the complete Minkowski spacetime we need three more sets of Rindler coordinates,
\begin{equation}\label{changeII}
ct=-\xi \sinh\left(\frac{a\tau}{c}\right),\qquad x=-\xi\cosh\left(\frac{a\tau}{c}\right),
\end{equation}
for region II, corresponding to an observer decelerating with respect to the Minkowskian origin, and
\begin{equation}\label{changeFP}
ct=\pm\xi \cosh\left(\frac{a\tau}{c}\right),\qquad x=\pm\xi\sinh\left(\frac{a\tau}{c}\right)
\end{equation}
for regions F and P.

Notice that for both relevant regions (I and II), the coordinates $(\xi,\tau)$ take values in the whole domain $(-\infty,+\infty)$. Therefore, if there is a quantum field defined in this flat spacetime, it is possible to carry out completely independent canonical field quantzsations of the field in regions I and II.

For simplicity, consider first an inertial observer in a flat spacetime (Alice) whose proper coordinates are the Minkowskian coordinates $(x,t)$. She wants to build a quantum field theory for a free massless scalar field.

She first needs to find an orthonormal basis of solutions of the free massless Klein-Gordon equation in Minkowski coordinates. As a reasonable possible choice, she can always use the positive energy plane wave solutions of the Klein-Gordon equation (and their complex conjugates) to build a complete set of solutions.

In this fashion the states $\ket{1_{\hat \omega}}_\textmod{M}=a^\dagger_{\hat
\omega,\textmod{M}}\ket{0}_\textmod{M}$ are free massless scalar field
modes, in other words, solutions of positive frequency $\hat\omega$
(with respect to the Minkowski timelike Killing vector $\partial_{
t}$) of the free Klein-Gordon equation:
\begin{eqnarray}\label{modmin}
\ket{1_{\hat\omega}}_\textmod{M}&\equiv u_{\hat\omega}^\textmod{M}\propto\frac{1}{\sqrt{2{\hat \omega}}}e^{-i {\hat \omega} t},
\end{eqnarray}
where only the time dependence has been made explicit. The label $\textmod{M}$ just means that these states are expressed in the Minkowskian Fock space basis.

The field expanded in terms of these modes takes the usual form 
\begin{equation}\label{minkexp12}
\phi=\sum_i \left(a_{\hat\omega_i,\textmod{M}}u_{\hat\omega_i}^\textmod{M}+a_{\hat\omega_i,\textmod{M}}^\dagger u_{\hat\omega_i}^{\textmod{M}*}\right),
\end{equation}
where the sum is shorthand notation for an integral over a suitable measure of field mode frequencies. Here, we have eliminated redundant notation and M denotes that $u_{\hat\omega_i}^\textmod{M}$ and $a_{\hat\omega_i,\textmod{M}}$ are Minkowskian modes and operators.

An accelerated observer can also carry out a canonical quantization procedure in his own right, and thus define his own vacuum and excited states of the field. Actually, for constantly accelerated observers, there are two natural vacuum states associated
with the positive frequency modes as defined by them with support in regions $\textmod{I}$ and $\textmod{II}$
of the flat spacetime. These are $\ket{0}_\textmod{I}$ and
$\ket{0}_{\textmod{II}}$, and subsequently we can define the field
excitations for Rindler observers with proper coordinates $(\xi,\tau)$ as
\begin{eqnarray}\label{modrin}
\ket{1_\omega}_\textmod{I}&=a^\dagger_{\omega,\textmod{I}}\ket{0}_\textmod{I}
\equiv u_\omega^\textmod{I}\propto\frac{1}{\sqrt{2\omega}}e^{-i\omega  \tau},\nonumber\\
\ket{1_\omega}_{\textmod{II}}&=a^\dagger_{\omega,\textmod{II}}\ket{0}_\textmod{II}
\equiv u_\omega^\textmod{II}\propto\frac{1}{\sqrt{2\omega}}e^{i\omega  \tau}.
\end{eqnarray}
These modes are related by a spacetime reflection and only have support in regions I and II of the spacetime respectively. Despite being limited to the left and right wedges, these modes nevertheless suffice as a complete set of solutions of the Klein-Gordon equation in Rindler coordinates because they are complete on a Cauchy hypersurface (e.g.,~$t=0$). Therefore, we can now expand the field \eqref{minkexp12} in terms of these modes:
\begin{equation}\label{rindexp1}
\phi=\sum_i \left(a_{\omega_i,\textmod{I}}u_{\omega_i}^\textmod{I}+a_{\omega_i,\textmod{I}}^\dagger u_{\omega_i}^{\textmod{I}*}+a_{\omega_i,\textmod{II}}u_{\omega_i}^\textmod{II}+a_{\omega_i,\textmod{II}}^\dagger u_{\omega_i}^{\textmod{II}*}\right).
\end{equation}

Expressions \eqref{minkexp12} and \eqref{rindexp1} are exactly equal and therefore Minkowskian modes can be expressed as function of Rindler modes by means of the Klein-Gordon inner product,
 \begin{equation}\label{KGscc}
(u_j,u_k)=-\ii\int \text{d}\Sigma\, n^\mu\left(u_j\partial_\mu u_k^*-u_k^*\partial_\mu u_j\right),
 \end{equation}
where $d\Sigma$ is the volume element and $n^\mu$ is a future directed timelike unit vector that is orthogonal to $\Sigma$.  Because the left and right Rindler modes together are complete, we can expand $u_{\hat\omega_j}^\textmod{M}$ in terms of them:
\begin{equation}\label{minkowskirindlerexp}
u_{\hat\omega_j}^\textmod{M}=\sum_i\left(\alpha^{\textmod{I}}_{ji}u_{\omega_i}^\textmod{I}+
\beta^{\textmod{II}}_{ji}u_{\omega_i}^\textmod{II*}+\alpha^{\textmod{II}}_{ji}
u_{\omega_i}^\textmod{II}+\beta^{\textmod{I}}_{ji}u_{\omega_i}^\textmod{I*}\right).
\end{equation}
Taking the Klein-Gordon inner product of this with $u_{\omega_j}^\Sigma$ and also with $u_{\omega_j}^{\Sigma*}$, with $\Sigma \in \{\text{I}, \text{II}\}$, %
allows us to extract the Bogoliubov coefficients,
\begin{equation}\label{bogo11}
\alpha^{\Sigma}_{ij}=\left(u_{\hat\omega_i}^{\textmod{M}},u_{\omega_j}^\Sigma\right),
\qquad\beta^{\Sigma}_{ij}=-\left(u_{\hat\omega_i}^{\textmod{M}},u_{\omega_j}^{\Sigma*}\right),
\end{equation}
where we have used the properties of the Klein-Gordon inner product, $(u_j,u_k)=\delta_{jk}=-(u_j^*,u_k^*)$ and $(u_j,u_k^*)=0$.

Now, we would like to know how the creation and annihilation operators in the Minkowski basis are related to operators in the two Rindler bases. Since we know that $a_{\hat\omega_i,\textmod{M}}=(\phi,u_{\hat\omega_i}^\textmod{M})$, if we write $\phi$ in Rindler basis \eqref{rindexp1} we can readily obtain
\begin{equation}\label{temporary1}
a_{\hat\omega_i,\textmod{M}}=\sum_j\left[(u_{\omega_j}^\textmod{I},u_{\hat\omega_i}^\textmod{M})
a^{\phantom{\dagger}}_{\omega_j,\textmod{I}}+(u_{\omega_j}^{\textmod{I}*},u_{\hat\omega_i}^\textmod{M})
a^\dagger_{\omega_j,\textmod{I}}+(u_{\omega_j}^\textmod{II},u_{\hat\omega_i}^\textmod{M})
a^{\phantom{\dagger}}_{\omega_j,\textmod{II}}+(u_{\omega_j}^{\textmod{II}*},u_{\hat\omega_i}^\textmod{M})
a^\dagger_{\omega_j,\textmod{II}}\right).
\end{equation}
which, using again the properties of the KG product,
\begin{equation}\label{KGproperties}(u_1,u_2)=(u_2,u_1)^*\,, \qquad (u_1^*,u_2^*)=-(u_2,u_1)\,,\end{equation}
allows us to readily write \eqref{temporary1} in terms of the Bogoliubov coefficients \eqref{bogo11} as 
\begin{equation}\label{inmink}
a_{\hat\omega_i,\textmod{M}}=\sum_j\left(\alpha^{\textmod{I}*}_{ij}
a^{\phantom{\dagger}}_{\omega_j,\textmod{I}}-\beta^{\textmod{I}*}_{ij}
a^\dagger_{\omega_j,\textmod{I}}+\alpha^{\textmod{II}*}_{ij}
a^{\phantom{\dagger}}_{\omega_j,\textmod{II}}-\beta^{\textmod{II}*}_{ij}
a^\dagger_{\omega_j,\textmod{II}}\right).
\end{equation}
 Completely  analogous reasoning can be followed for the case of a Dirac field, with some subtleties that are partially discussed in~\cite{Edu2,Edu4,beyond}.  

To obtain the form of the Minkowskian vacuum in the Rindler basis, we have to demand that \eqref{inmink} (the Minkowskian annihilation operator) annihilates the vacuum for all $\omega_i$. This in turn implies that the right-hand-side of \eqref{inmink} has to annihilate the Minkowski vacuum in the Rindler basis. It is straightforward to see that the Rindler basis representation of the Minkowskian vacuum has to be a Gaussian state \cite{NavarroSalas}. Assuming this ansatz and demanding that the r.h.s of \eqref{inmink} annihilates the Minkowski vacuum we obtain an infinte number of recursive equations that, together with the normalization condition, allows us to find the exact form of the state in terms of the Bogoliubov coefficients. These coefficientes will  be continuous linear combinations of Rindler annihilation and creation operators for all Rindler frequencies. All this imposes that the vacuum state has to be a tensor product of two-mode squeezed states for every Rindler frequency \cite{Kerr,colapse,NavarroSalas}.

After all this algebra, one finds that the Minkowskian basis can be expressed as a two mode squeezed state in the Rindler basis~\cite{Takagi,NavarroSalas}. Namely, for the scalar case considered in this introduction
\begin{equation}\label{scalarvacuum1}
\ket{0}_{\textmod{M}}=\bigotimes_{\omega}\frac{1}{\cosh r_{\textmod{b},\omega} }\sum_{n=0}^\infty \tanh^n r_{\textmod{b},\omega} \ket{n}_{\omega,\textmod{I}}\ket{n}_{\omega,\textmod{II}}.
\end{equation}
where\footnote{The label b stands for `bosonic', this parameter has a different definition for fermionic and bosonic  fields~\cite{Edu2}.}
\begin{equation}\label{rbos1}
r_{\textmod{b},\omega}=\textmod{atanh} \left[\exp\left(-\frac{\pi c\omega }{a}\right)\right].
\end{equation}

\subsection*{The Unruh effect}\label{tue}

In the  1970s,  Fulling, Davies and Unruh realised that the impossibility  of covering the whole of Minkowski spacetime with only one set of Rindler coordinates has strong implications when accelerated and inertial observers describe states of a quantum field. Namely, the description of the vacuum state of the field in the inertial basis as seen by accelerated observers has a non-zero particle content. 

In very plain words, the Unruh effect is the fact that while inertial observers `see' the vacuum state of the field, an accelerated observer would `see' a thermal bath whose temperature is proportional to his acceleration.

Different approaches to this well-known effect can be found in multiple textbooks (let us cite~\cite{Wald2,Birrell,NavarroSalas} as a token). Nowadays, the use of particle detectors is the preferred way to show the Unruh effect~\cite{Crispino}, this allows us to escape the common caveats of  derivations based on the direct analysis of the quantum field.

However, in this section, we will provide a different rather simple derivation of the effect in a way that will be useful in order to clearly present a feature of spacetime with horizons which turns out to be relevant when it comes to study entanglement.

Let us consider that an inertial observer, Alice, is observing the vacuum state of a scalar field. Now assume that an accelerated observer, called Rob,  wants to describe the same quantum field state by means of his proper Fock basis. One way to find Rob's description of the field is to find the change of basis from the Fock basis build from solutions to the Klein-Gordon equation in Minkowskian coordinates \eqref{modmin} to the Fock basis build from solutions of the KG equations in Rindler coordinates \eqref{modrin}. This gives us equation \eqref{scalarvacuum1} which we presented in the section above.

Of course, \eqref{scalarvacuum1} is a pure state. However, the eternally accelerated observer is restricted to region I (or II) of spacetime due to the presence of an acceleration horizon (as shown in Figure \ref{rin2}), and, since both regions are causally disconnected, Rob has no access to any degrees of freedom which have support in the opposite spacetime wedge, which have to be traced out from his quantum state description. This means that the quantum state accessible for Rob is no longer pure,
\begin{equation}
\rho_{\textmod{R},\omega}=\tr_{\textmod{II}}\left(\proj{0_\omega}{0_\omega}\right)=\sum_{k}\bra{k}_{\textmod{II}}\ket{0_\omega}_\textmod{M}\bra{0_\omega}_\textmod{M}\ket{k}_{\textmod{II}}.
\end{equation}
Substituting $\ket{0_\omega}$ by its Rindler basis expression \eqref{scalarvacuum1} we obtain
\begin{equation}\label{thermal}
\rho_{\textmod{R},\omega}=\frac{1}{\cosh^2 r_{\textmod{b},\omega}}\sum_n\tanh^{2n}r_{\textmod{b},\omega}\ket{n_\omega}_\textmod{I}\bra{n_\omega}_\textmod{I},
\end{equation}
which is a thermal state.

Indeed, if we compute the particle counting statistics that the accelerated observer would see we obtain
\begin{equation}
\langle N_{\omega,\textmod{R}}\rangle=\tr_{\textmod{I}}\left(\rho_{\textmod{R},\omega}\,a^\dagger_{\omega,\textmod{I}}a_{\omega,\textmod{I}}\right)=\frac{1}{e^{2\pi c/\omega a}-1},
\end{equation}
which is a Bose-Einstein distribution with temperature
\begin{equation}
T_\textmod{U}=\frac{\hbar a}{2\pi K_\textmod{B}},
\end{equation}
 the so-called Unruh temperature.

Rob observes a thermal state\footnote{Note that thermal noise is only observed in the 1+1 dimensional case. In higher dimension Rob would observe a noisy distribution, very similar to a thermal one, but with different prefactors~\cite{Takagi}. This is called the Rindler noise.} because he cannot see modes with support in region II due to the presence of an acceleration horizon.  Curiously, the essence of this effect persists even if there is, strictly speaking, no horizon for the observer---for instance, in the presence of a reflecting boundary condition at the origin~\cite{Rovelli:2012jm}. The presence of a horizon as described above, however, allows for an exactly thermal response.

\subsection{Bogoliubov transformations and particle production in asymptotically stationary spacetimes and in cosmology}
\label{nonstaint}

The analysis carried out in the previous subsection  assumed that the spacetime is stationatry and therefore  possesses a global time isometry. In general, this treatment would not be possible in the absence of a global timelike Killing vector. However, there are some scenarios in which the spacetime is not stationary but possesses stationary asymptotic regions. This is the case, for example, of some models of expansion of the universe~\cite{dun1} or the stellar collapse and formation of black holes~\cite{NavarroSalas}. For the sake of completeness will summarise in this subsection the analysis detailed in \cite{MartinMartinez2012}.

Consider a spacetime that has asymptotically stationary regions in the far past and in the far future. We denote them `in' and `out' respectively. In these regions it is possible to ascribe a particle interpretation to the solutions of the field equations. Namely, we can build two different sets of solutions of the field, the first one~$\{u_{\hat{\bm{k}}_j}^{\mathrm{in}}\}$ consisting of modes with positive frequency~$\hat\omega_j$ with respect to comoving time in the asymptotic past. The second set~$\{u_{{\bm{k}}_j}^{\mathrm{out}}\}$ consisting of modes with positive frequency~$\omega_j$ with respect to comoving time in the asymptotic future. In this fashion we can expand the quantum field in terms of both sets of solutions
\begin{equation}\label{minkexp1}
\Phi=\sum_i \left(a_{\hat{\bm{k}}_i,\mathrm{in}}u_{\hat{\bm{k}}_i}^{\mathrm{in}}+a_{\hat{\bm{k}}_i,\mathrm{in}}^\dagger u_{\hat{\bm{k}}_i}^{\mathrm{in}*}\right)=\sum_i \left(a_{{\bm{k}}_i,\mathrm{out}}u_{{\bm{k}}_i}^{\mathrm{out}}+a_{{\bm{k}}_i,\mathrm{out}}^\dagger u_{{\bm{k}}_i}^{\mathrm{out}*}\right)\,.
\end{equation}
Now, since both set of modes are complete, we can expand the particle operators associated with one basis in terms of operators of the other basis~\cite{MartinMartinez2012,NavarroSalas}:
\begin{eqnarray}
\label{uno1}a_{\hat{\bm{k}}_i,\mathrm{in}}&=\sum_j \left(\alpha_{ji} a_{{\bm{k}}_j,\mathrm{out}}+\beta_{ji}^* a_{{\bm{k}}_j,\mathrm{out}}^\dagger\right),\\
\label{uno2}a_{{\bm{k}}_i,\mathrm{out}}&=\sum_j \left(\alpha^*_{ij} a_{\hat{\bm{k}}_j,\mathrm{in}}-\beta_{ij}^* a_{\hat{\bm{k}}_j,\mathrm{in}}^\dagger\right).
\end{eqnarray}
Now let us consider the vacuum state in the asymptotic past region $\ket0_{\mathrm{in}}$, which is annihilated by $a_{\hat{\bm{k}}_i,\mathrm{in}}\ \forall\hat\omega_i$. We would like to know the form of the state $\ket0_{\mathrm{in}}$ in the basis of modes in the asymptotic future. To compute this we use the fact that  $a_{\hat{\bm{k}}_i,\mathrm{in}}\ket0_{\mathrm{in}}=0$. If we substitute $a_{\hat{\bm{k}}_i,\mathrm{in}}$ in terms of `out' operators using Eq.~\eqref{uno1}, we obtain
\begin{equation}\label{condit}
\sum_j \left(\alpha_{ji} a_{{\bm{k}}_j,\mathrm{out}}+\beta_{ji}^* a_{{\bm{k}}_j,\mathrm{out}}^\dagger\right)\ket{0}_{\mathrm{in}}=0\,.
\end{equation}

It is relatively straightforward to prove  (See~\cite{MartinMartinez2012}) that this condition implies that the `in' vacuum can be written in terms of `out' modes as
\begin{equation}\label{dynoresult}
\ket{0}_{\mathrm{in}}=C\exp\left(-\frac12\sum_{ijk}\beta^*_{ik}(\alpha^{-1})_{kj}a_{{\bm{k}}_{i},\mathrm{out}}^\dagger a_{{\bm{k}}_{j},\mathrm{out}}^\dagger\right)\ket{0}_{\mathrm{out}}\,.
\end{equation}
This is a Gaussian state and $C$ is given by the normalization condition. Notice that this state is, in general, not separable. This means that, depending on $\alpha_{ij}$ and $\beta_{ij}$, the final state of the field can possess nontrivial quantum correlations.

Similar to what we did in the small subsection about the Unruh effect,  we could compute here the expectation value of the number operator in the asymptotic future with respect to the `in' vacuum, finding
\begin{equation}\label{production}
\langle N^{\mathrm{out}}_{{\bm{k}}_{j}}\rangle_{\mathrm{in}} = {}_{\mathrm{in}}\!\bra{0}a_{{\bm{k}}_{j},\mathrm{out}}^\dagger a_{{\bm{k}}_{j},\mathrm{out}}\ket{0}_{\mathrm{in}}=\sum_i \left|\beta_{ij}\right|^2.
\end{equation}
This implies that if $\beta_{ij}$ is different from zero, one would observe particle production as a consequence of the expansion.

Now, as a relevant particular case, let us consider a universe undergoing a homogeneous and isotropic expansion, which is very well described in terms of the Friedmann-Lema\^itre-Robertson-Walker~(FLRW) metric, 
\begin{equation}\label{FLRWm1}
ds^2=dt^2-[a(t)]^2 d\Sigma^2\,,
\end{equation}
where $d\Sigma$ is an element of a 3-dimensional space of uniform curvature, either elliptical, hyperbolic, or Euclidean, this is,
\begin{equation}
ds^2=dt^2-[a(t)]^2 \left(\frac{dr^2}{1-kr^2}+r^2d\Omega^2\right)\,,
\end{equation}
where $d\Omega$ is the line element in the unit sphere and $k$ characterises the curvature of the space (negative, zero, or positive for hyperbolic, Euclidean, or elliptical, respectively). We will consider that the spatial geometry is Euclidean ($k=0$), so we can write this metric as
\begin{equation}\label{FLRWm}
ds^2=dt^2-[a(t)]^2 (dx^2+dy^2+dz^2)\,.
\end{equation} Even if the spatial geometry of the universe is not exactly Euclidean, all evidence seems to indicate that the spatial curvature is close enough to zero to place the radius at approximately the horizon of the observable universe or beyond, so this consideration seems reasonable.

As discussed above, when a non-stationary spacetime possesses asymptotically stationary regions, it is possible to compute expectaiton values of observables in the asymptotic future by knowing the field state in the asymptotic past, which enables us to compute the field quanta creation due to the gravitational interaction. This is the case for the FLRW spacetime,
 Eq.~\eqref{FLRWm}, if we impose the additional conditions
\begin{equation}\label{condasy}
a(+\infty)\rightarrow \mathrm{const.}\,, \qquad  a(-\infty)\rightarrow \mathrm{const.}\,,
\end{equation}
Although measurements of distant supernovae have demonstrated that our universe is undergoing an accelerating expansion, to good degree of approximation, certain models fulfilling Eq.~\eqref{condasy} can serve as approximations  to some inflationary scenarios. Concretely, they can be rough approximations to a period of slow expansion, then a period of very fast expansion, and then again a period of slow expansion.

As can be seen from Eq.~\eqref{uno2}, computing the analytic form of the particle operators in the asymptotic future in terms of the operators in the asymptotic past is generally not trivial. There are, however, some particular solvable toy models that allow for the study of fundamental phenomena in this kind of spacetime. These include some scalar models~\cite{Birrell} and spin-$\frac 1 2$ fermionic models~\cite{dun1}. To illustrate this, let us revisit the scalar field case within an exactly solvable model in 1+1 dimensions analyzed in~\cite{MartinMartinez2012}.

Let us consider the FLRW metric Eq.~\eqref{FLRWm} and rewrite it in terms of the conformal time coordinate
\begin{equation}\label{conFLRW}
\eta=\int_0^t \frac{\mathrm{d}\tau}{a(\tau)}\,.
\end{equation}
This yields,
\begin{equation}\label{MconFLRW}
ds^2=[a(\eta)]^2(d\eta^2- dx^2)\,.
\end{equation}
Following the usual notation, we define $C(\eta)=[a(\eta)]^2$. We now  assume the following specific form for the conformal factor:
\begin{equation}\label{factorbos}
C(\eta)=1+\epsilon \tanh(\rho\eta)\,,
\end{equation}
where $\epsilon$ and $\rho$ are positive real parameters characterising the total volume and rapidity of the expansion, respectively. Imposing this particular form for the conformal factor makes it simple to tackle the problem analytically, while at the same time allows us to study the fundamental behaviour of the particle creation phenomenon as a function of the rapidity of the expansion and its total volume.

It is interesting to note that for large $\epsilon$ the FLRW metric with this conformal factor behaves, for $0<t\ll \rho^{-1}$, like the radiation-dominated Friedman universe, with an exponentially fast approach to asymptotic flatness for $t\ll \rho^{-1}$. As mentioned above, we can consider this as a convenient way to approximate a spacetime that is asymptotically flat in the past and future but undergoes rapid inflation in between. The asymptotic flatness at each end allows us to define different but physically meaningful particle states for the asymptotic past and future. Note that this example is analytically solvable only for the scalar field. The same $C(\eta)$ does not provide an analytically solvable model in the fermionic case, which will be discussed in Sec.~\ref{subsec:bosfermfieldent}.

We label the mode solutions to the Klein-Gordon equation in the asymptotic past as~$u^{\mathrm{in}}_k(x,\eta)$ and those in the asymptotic future as~$u^{\mathrm{out}}_k(x,\eta)$. The two are connected by a Bogoliubov transformation that only mixes modes of the same~$k$:
\begin{equation}\label{modolibov}
u^{\mathrm{in}}_k(x,\eta)=\alpha_k  u^{\mathrm{out}}_k(x,\eta)+\beta_k u^{{\mathrm{out}*}}_{-k}(x,\eta)\,.
\end{equation}
The reader is referred to Refs.~\cite{Birrell1982,MartinMartinez2012} for details of this derivation.\footnote{Note that Eq.~(44) in Ref.~\cite{MartinMartinez2012}, which corresponds to Eq.~\eqref{modolibov} above, contains a typo: what should be $u^{{\mathrm{out}*}}_{-k}$ appears as $u^{{\mathrm{in}*}}_{-k}$ in that work.}

The fact that the Bogoliubov transformation only mixes modes of the same momentum is not related to the particular expansion model choice in Eq.~\eqref{factorbos}, nor to the dimension of the spacetime considered. Rather, it is a consequence only of the conformal symmetry (up to mass terms in the minimal coupling scenario) of the theory and the conformal flatness of the metric. This conformal equivalence relates the equation of motion in the FLRW scenario to its flat-spacetime form. This means that the general form of the Bogoliubov transformation Eq.~\eqref{modolibov} is valid for any FLRW universe with sufficiently well-behaved scaling factor and minimally coupled field, where Eq.~\eqref{condasy} is fulfilled~\cite{dun1}.

From Eq.~\eqref{production} we see that the average number of particles in the asymptotic future when the field was prepared in the vacuum state in the past is
\begin{equation}
\langle N^{\mathrm{out}}_{k}\rangle_{\mathrm{in}}=|\beta_k|^2.
\end{equation}
This can be interpreted as the creation of quanta in the mode $k$ of the field as a consequence of the spacetime expansion. Notice that when $m=0$, $\omega_{\mathrm{in}}=\omega_{\mathrm{out}}$, which means that $|\beta_k|=0$, and no particles are present in the asymptotic future. This is because, in the particular case of two spacetime dimensions, for $m=0$, the theory is also conformally invariant. We will analyze this case in more detail.

From a general perspective, Hilbert-Einstein action of a scalar field in a curved spacetime  of dimension~$D$, takes the form
\begin{equation}\label{action}
\mathcal{S}=\int \mathrm{d}^D x \frac{\sqrt{|g|}}{2}\left[g^{\mu\nu}\partial_\mu \Phi\partial_\nu \Phi-(m^2+\xi R)\Phi^2 \right]\,,
\end{equation}
whose associated equation of motion is
\begin{equation}\label{field}
\left(\da + m^2+\xi R \right)\Phi=0\,,
\end{equation}
 $\xi$ is the coupling of the field to the Ricci curvature scalar~$R$. The case where $\xi=0$ is called minimal coupling. There is a particular value of this coupling strength which is of great interest in cosmological scenarios given that the FLRW metric is conformally flat:  \emph{conformal coupling.}.

If we require the action in Eq.~\eqref{action} to be invariant (except for an irrelevant boundary term) under a conformal transformation $g'_{\mu\nu}=[\Omega(x)]^2g_{\mu\nu}$, we find that there are three contributions that may destroy this invariance: the mass term, the derivatives in the kinetic term (if $\Omega$ is not constant), and the coupling to the curvature term~\cite{MartinMartinez2012}. In the massless case, we can always choose the coupling to the curvature such that the action becomes conformally invariant---namely,
\begin{equation}
\xi=\frac{(D-2)}{4(D-1)}\,,
\end{equation}
Where $D$ is the spacetime dimension. This is the so-called conformal coupling. In two spacetime dimensions, we see that if $m=0$ the minimal coupling $\xi=0$ gives conformal invariance of the action, so it coincides exactly with the conformal coupling. In contrast, in four spacetime dimensions, we require $m=0$  and $\xi=1/6$ to have a conformally invariant action. 

Under conformal symmetry of the action, if $\Phi'$ is a solution of the conformally transformed equation, then $\Phi=\Omega\Phi'$ will be a solution of the original equation~\cite{MartinMartinez2012}.

The FLRW metric from Eq.~\eqref{FLRWm} happens to be conformally flat, as it is explicitly displayed when written in terms of the conformal time $\eta$
\begin{equation}
ds^2=[a(\eta)]^2(d\eta^2- dx^2-dy^2-dz^2)\,.
\end{equation}
This allows us to use the conformal symmetry to find the form of the solutions to the KG in the FRLW background:
\begin{equation} 
u_{\bm k}\propto [a(\eta)]^{-1}\exp[\ii(\bm k \cdot \bm x- |\bm k| \eta)]=[a(\eta)]^{-1}\exp\left[\ii\left(\bm k \cdot \bm x- \int_0^t \omega(\tau)\, \mathrm{d}\tau \right)\right]\,,
\end{equation}
where we have defined $\omega(t)=|\bm k|/a(t)$. $u_{\bm k}$ are positive frequency solutions at early times, so the field admits the expansion
\begin{equation}
\Phi=\sum_i \left(a_{\hat{\bm{k}}_i,\mathrm{in}}u_{\hat{\bm{k}}_i}+a_{\hat{\bm{k}}_i,\mathrm{in}}^\dagger u_{\hat{\bm{k}}_i}^{*}\right)\,.
\end{equation}
But $u_{\bm k}$ are also positive-frequency solutions at late times, so the field also admits the expansion
\begin{equation}
\Phi=\sum_i \left(a_{{\bm{k}}_i,\mathrm{out}}u_{{\bm{k}}_i}+a_{{\bm{k}}_i,\mathrm{out}}^\dagger u_{{\bm{k}}_i}^{*}\right)\,.
\end{equation}
In this case, the Bogoliubov transformation is trivial, and $a_{{\bm{k}}_i,\mathrm{out}}=a_{{\bm{k}}_i,\mathrm{in}}$, so that the early- and late-time vacua are the same. 

In conclusion, under conformal symmetry the concept of particle is well-defined at all times, and there is therefore a natural choice for the vacuum state~\cite{Parker1,Parker2,Parker3}, which is called the {\it conformal vacuum}~\cite{Birrell,Parker1}. As an obvious consequence, the Bogoliubov coefficients $\beta_{\bm k}=0$ for all~$\bm k$, so no particles are created in the massless conformally coupled field case. Hence, if the field is prepared in the conformal vacuum in the asymptotic past, the expansion will not create any particles in the asymptotic future either. 

Nevertheless, it would not be completely correct to assert that there is no particle creation due to the expansion of the universe insofar as we suscribe to Unruh's definition of particle as ``what particle detectors click for''. As we will see in the next section, a comoving particle detector can nevertheless `click' in this conformally symmetric scenario.

\section{Particle detectors in curved spacetimes: perturbative and non-perturbative analysis}

\subsection{The Unruh-DeWitt (UDW) model of particle detectors}

Particle detector models, such as the Unruh-Dewitt model, have been used since their inception to quantify the particle content in a given state of a quantum field as seen by a localised observer \cite{Unruh1976,DeWitt,Birrell,Wald2,Crispino};
to harvest the entanglement of the vacuum state of a quantum field \cite{Reznik2005,farming}; in metrology settings \cite{Dragan:2011zz,Quantumthermometer,Quantumthermometer2,marvy1,marvy2}; to analyze the decoherence effects of relativistic trajectories \cite{lmt10,matsako}, to propose schemes of universal quantum computing via relativistic motion  \cite{AasenPRL,Chris} and to set up scenarios of quantum communication  in the relativistic limit \cite{Robort}. 
Considering a localised point-like quantum system coupled to the field as a particle detector has the advantage that we can associate every detector with a particular `observer' who is moving through the spacetime.

Particle detector models were introduced by Unruh~\cite{Unruh1976} to give a more physical approach to the results by Fulling~\cite{Fulling1973} and Davies~\cite{Davies1975} on the effects of acceleration on the observed state of a quantum field for field modes. Unruh showed that the effect is real and felt by a local particle detector~\cite{Unruh1976}. He used two different models: (1)~a particle in a box coupled to a scalar field and (2)~two quantum fields, one heavier than the other, whose interaction is mediated by the external field. The purpose of these two alternatives was to show that the results obtained for the semiclassical model of a particle in a box displays the same phenomenology as a fully second-quantised model. Shortly thereafter, DeWitt~\cite{DeWitt:1979tl} discussed that only the two lowest energy levels of the particle in a box were really necessary to see the relevant phenomenology, and nowadays a qubit is most commonly used as the system serving as the semiclassical detector in what is known as the UDW model. This model, whose applicability relies heavily on the existence of a perturbative approach,  will be briefly reviewed in Sec.~\ref{qubitUDW}. More recently it was shown that in the case of free bosonic fields, we can obtain nonperturbative solutions for the behavior of particle detectors by adding back additional levels to the  detector---but as a harmonic oscillator instead of the original box~\cite{Unruh1976}. This lets us model the evolution as symplectic transformations on phase space, allowing for nonperturbative probing of detector-field interactions. This powerful new technique is reviewed in Sec,~\ref{harmoUDW}.

\subsection{Qubit-based Unruh-DeWitt detectors}\label{qubitUDW}

The detector model consists of a two-level quantum system with a scalar field. In its simplest form, this is a point-like particle detector whose interaction Hamiltonian with a scalar field~$\phi(x)$ is
\begin{equation}
\label{eq:UdW}
H_I=\lambda \chi(\tau)\,\mu(\tau)\,\phi[x(\tau)]\,,
\end{equation}
where $\lambda$ is the coupling strength, $\tau$ is the proper time of the detector, $\chi(\tau)$ is some appropriate switching function describing the time-dependent coupling of the detector to the field, $\mu(\tau)$ is the monopole moment of the detector, and $\phi[x(\tau)]$ is the field operator evaluated along the worldline of the detector. The detector being a two-level quantum system, the monopole momentum operator can be written in terms of qubit ladder operators $\sigma^+$ and  $ \sigma^-$. We can choose to expand the field operators in terms of positive- and negative-frequency solutions of the Klein-Gordon equation in a given quantization frame $(x,t)$: $u_k(x,t)$ and $u_k^*(x,t)$, yielding the interaction picture Hamiltonian
\begin{eqnarray}
\label{Hamilera}
H_{I}=\lambda\,\chi(\tau) \left(\sigma^+\ee^{\ii \Omega \tau}+\sigma^-\ee^{-\ii\Omega \tau}\right)  \int\mathrm{d} k  \Big(a^\dagger_{k} u^*_k[x(\tau),t(\tau)]+a_{k} u_k[x(\tau),t(\tau)]\Big)\,,
\end{eqnarray}
where $\Omega$ is the energy gap between the ground and excited state of the detector.

The Unruh-Dewitt model is, by construction, simple but it encompasses the fundamental features of the light-matter interaction when  transitions between two atomic levels where no orbital angular momentum exchange occurs  \cite{Wavepackets,Alvaro}. 
A series of further approximations (single-mode and rotating-wave) carried out on the UDW detector model yields the Jaynes-Cummings model, which is commonplace in quantum optics as a phenomenological model of light-matter interaction \cite{Scullybook}. 
Additionally, the UDW detector is a powerful effective model in superconducting-circuit QED \cite{supercond,Wallraff04}. 

It is well-known that an accelerated Unruh-DeWitt detector has a thermal response~\cite{Birrell}, a result that is very robust under small perturbations of the trajectory or imposition of finite boundary conditions \cite{Wilson,LuisBarbado,Manno}. In comsological contexts, it is also well-known that if we consider a single UDW particle detector in a de~Sitter universe coupled to a quantum scalar field in the conformal vacuum~\cite{Birrell1982}, at the first order in perturbation theory and for long interaction times $\chi(\tau)\approx1$, the response function of such detector is~\cite{Birrell}
\begin{equation}
F(E)=\frac{1}{4\pi^2}\int \mathrm{d}\eta\int\mathrm{d}\eta'\,\frac{\ee^{-\ii E\int_{\eta'}^\eta\sqrt{a(\eta'')}\mathrm{d}\eta''}}{(\eta-\eta'-\ii\epsilon)^2}\,,
\end{equation}
which is nonzero in general. Hence, a comoving detector will click even if the state of the field is the conformal vacuum. A paradigmatic example of this is the de~Sitter universe, in which a comoving detector gives a thermal response~\cite{GibHawking}. The characteristic temperature of the detected particle statistics is known as the Gibbons-Hawking temperature and is proportional to the expansion rate.  Notice that despite the lack of particle creation (in the asymptotic sense) in the conformal vacuum, particle detectors nevertheless click due to the transient effects of expansion. 

\subsection{Non-perturbative particle detectors}\label{harmoUDW}

Although the Unruh-DeWitt model has been very successful in dealing with problems related to the particle content of  quantum fields and the harvesting of field entanglement, its main shortcoming is that almost all the work based on such detectors so far has been limited to perturbation theory. This limitation rendered the model unsuitable to study problems in which a perturbative expansion is not a good approximation. These include strong coupling, long times and high-average-energy exchange processes.

Recently, two groups proposed modelling a detector as a quantum harmonic oscillator rather than as a qubit~\cite{Brown,Lee}. In other words, the authors simply replace two energy levels with infinitely many evenly-spaced levels. The use of such model  has been proposed before in other contexts~\cite{UnrhZurek,HuMatacz,MassarSpindel,BeiLok,BerryPh,Ivette}. Considering this detector instead of a qubit detector is not a downgrade of the model. Qubits are, in many cases, just approximations to systems with many more levels. For example,  most symmetric potentials in nature can be approximated by a harmonic potential for low energies, so a harmonic-oscillator detector can model a wide range of different detector systems, from atomic electromagnetic levels to the molecular vibrational spectrum. 

Although both Ref.~\cite{Brown} and Ref.~\cite{Lee} make use of the fact that Gaussian unitary operations act via symplectic transformations on phase space, the two approaches differ in the details of their implementation. In Ref.~\cite{Lee}, the symplectic dynamics are expanded in terms of the symplectic algebra, which results in an implementation that has a minimal number of terms and thus a minimal number of differential equations to solve. In Ref.~\cite{Brown}, on the other hand, the authors represent the dynamics using the full matrix representation of the symplectic transformation, which includes some redundancy in the representation of the dynamical objects---and thus more differential equations to solve. It has the distinct advantage, however,  that the equations to be solved are always linear, first-order, ordinary differential equations. In the rest of this section, we describe the techniques of Ref.~\cite{Brown} in detail since these techniques have already proven useful for practical calculations~\cite{Brown,farming,Wilson,Brown:2014uj}, and we refer the reader to Ref.~\cite{Lee} for details of the alternate method, which may yet prove to be advantageous in certain scenarios.

The idea of using harmonic oscillators in relativistic quantum field theory as particle detectors to obtain non-perturbative results was first explored by Bei-Lok Hu and collaborators, who reported interesting analytical results in \cite{BeiLok}. Along with its considerable technical accomplishments, this approach emphasised that the Unruh effect is not reliant on gravitational or geometrical arguments but can be understood as a dynamical effect insofar as it indicates how the quantum vacuum affects the response of a detector contingent on its motion. In general, a detector detects   field quanta with a nonthermal spectrum, where the degree of nonthermality is governed by the parameter that measures the deviation from uniform acceleration \cite{Raval:1996vt}.   However the practical scope of this approach remains to be seen---thus far
it has been limited to very concrete problems in relativistic quantum theory \cite{Lin:2008jj,Hu,Ostapchuk:2011ud}
due to their complexity and the number of assumptions and approximations required to obtain quantitative results. 

The main development in \cite{Brown} is the use of the symplectic formalism for Gaussian states and operations~\cite{Schumaker1} applied to generally time dependent multimode scenarios. In doing  so, the authors  prove that  the  quantum evolution of relativistic particle detectors coupled to quantum fields can be solved \emph{nonperturbatively}. Many of the scenarios of interest in relativistic quantum theory involve quadratic Hamiltonians and Gaussian states (vacuum, thermal states, squeezed states, etc.), making this formalism widely applicable.

In particular,  the authors show how this approach can be used to compute the full time evolution of systems of detectors coupled to quantum fields in cavity settings for
\begin{enumerate}
\item arbitrary time-dependent trajectories and spacetime backgrounds;
\item arbitrary quadratic, time-dependent interaction Hamiltonians and boundary conditions;
\item arbitrary Gaussian initial states of the field modes and detectors;
\item any number of detectors;
\end{enumerate}
all beyond the limitations of perturbation theory. With this formalism, it is possible to overcome causality violation problems of  single (or few) mode approximations for detectors undergoing general trajectories \cite{Fay,Robort}.  The results in \cite{Brown} are devoid of faster-than-light signalling, unlike previous results limited to single mode approaches \cite{Ivette}.

We would like to emphasise the importance of this non-perturbative approach:  The usual perturbative methods cannot analyze physically interesting scenarios as, for instance, ultra-strong coupling in circuit QED \cite{ultrastrong}, or the problem of thermalization of an accelerated Unruh-Dewitt detector \cite{Wilson}.

The limitation of this approach is that, in practice, one is forced to apply an infrared cutoff to the field.  However, an infrared cutoff naturally appears when studying quantum field theories in finite volumes (e.g., optical  cavities, periodic waveguides, etc.), and so this formalism enables us to non-perturbatively solve problems of quantum field theories in curved spacetimes inside cavities, a matter of great interest that has not been thoroughly explored to date. If a tabletop experiment in which relativistic quantum phenomena is to appear, discrete systems~\cite{Menicucci2010a} or superconducting circuits \cite{Sab1,PastFutPRL} have an edge on  experimental feasibility.

Let us briefly summarise the technique following \cite{Brown}: We write the most general X-X type Hamiltonian for an arbitrary number of detectors undergoing general trajectories with different proper times $\tau_j$ and with general time dependent couplings. We choose to take the quantization frame to correspond to a stationary inertial observer $t$. Writing the Hamiltonian that generates translations with repsect to the time parameter $t$ in the Heisenberg picture yields
\begin{equation}\label{eq:hamilto}
\nonumber \op H^{\text{H},t}=\sum_{n=1}^N \omega_n \op a_n^\dag \op a_n 
+\sum_{j=1}^M \frac{d\tau_j(t)}{dt}\Big[\Omega_j \op a_{d_j}^\dag \op a_{d_j}+\sum_{n=1}^N \lambda_{nj}(t)(\op a_{d_j}+\op a^\dag_{d_j})\big(\op a_n v_n[x_j(t)] +\op a^\dag_n v_n[x_j(t)] \big)\Big]
\end{equation}
where $x_j(t)$ is the trajectory of the $j$-th detector parametrised in terms of the global Minkowskian time $t$, and all operators are now understood to be in their Heisenberg representation. Working in this representation allows us to derive a simple, number-valued equation of motion that describes the full evolution of the detectors+field state.

Following \cite{Brown} we consider the Heisenberg quadrature operators~$(\op q_{d_j}(t), \op p_{d_j}(t))$ for each detector and $(\op q_n(t), \op p_n(t))$ for each field mode. The usual compact notation consists of stacking these operators on top of each other to form the following vector of operators, while omitting the explicit $t$-dependence for clarity:
\begin{equation}
\opvec x \coloneqq \left(\op q_{d_1},\dots,\op q_{d_M}, \op q_{1},\dots,\op q_{N},\op p_{d_1},\dots,\op p_{d_M},\op p_{1},\dots,\op p_N\right)^\tp\,,
\end{equation}
where
\begin{equation}
 \op q_i=\frac{1}{\sqrt{2}}\big(\op a_i+\op a_i^\dag\big),\quad \op p_i=\frac{\ii}{\sqrt{2}}\big(\op a_i^\dag-\op a_i\big)
\end{equation}
are respectively the Heisenberg picture canonical position and momentum of every single oscillator. Note that the transpose operation~$^\tp$ merely transposes the shape of an operator-valued vector; it does nothing to the operators themselves.

If we neglect phase-space displacements,  a Gaussian state~\cite{Weedbrook:2012fe} is fully characterised by a covariance matrix~$\mat \sigma$, the entries of which are
\begin{eqnarray}
	\sigma_{i j} \equiv \langle \op{x}_i \op{x}_j + \op{x}_j \op{x}_i \rangle - 2\langle \op{x}_i \rangle \langle \op{x}_j \rangle\,.
\end{eqnarray}
Although in general, displacements will be required for a full description of Gaussian states and evolution, if the initial state is zero-mean Gaussian state (such as, e.g., the ground state for the detectors and the vacuum, squeezed vacuum or  thermal state for the field), then the lack of linear terms in our Hamiltonian means that the state remains at zero mean at all times.

To compute time evolution, we utilise the fact that quadratic Hamiltonians preserve Gaussianity~\cite{Schumaker1}. This means that quadrature operators get mapped to linear combinations of quadrature operators:
\begin{eqnarray}
\label{eq:Heisenbergsymplectic}
	\opvec x' = \op U^\dag \opvec x \op U = \mat S \opvec x\,,
\end{eqnarray}
where $\mat S$ is a symplectic matrix of c-numbers that acts via matrix multiplication on $\opvec x$ as a vector, while $\op U$ is a unitary operator that acts on the individual operators within~$\opvec x$.  Although  in general there would be a phase-space displacement term, which would give $\opvec x' = \mat S \opvec x + \vec y$,  we are neglecting this as justified above. 

The symplectic nature of~$\mat S$ is guaranteed because the commutation relations must be preserved, giving rise to a symplectic form~$\mat \Omega$ to be preserved by the Heisenberg matrix action.    Using the notation of Ref.~\cite{Brown}, $\mat \Omega$ has elements $\Omega_{ij}=-i \big[\op{x}_i,\op{x}_j \big]$. That $\mat S$ is symplectic means that $\mat S \mat \Omega \mat S^\tp = \mat \Omega$, which follows from requiring that the new commutation relations must be fulfilled at all times: $[\mat S \opvec x, (\mat S \opvec x)^\tp] = i\mat \Omega$. (See Ref.~\cite{Brown,Menicucci2011} for more details.)

A general quadratic Hamiltonian generates a Gaussian unitary~$\op U(t)$
that is associated, by Eq.~\eqref{eq:Heisenbergsymplectic}, with a symplectic matrix~$\mat S(t)$, which satisfies
\begin{eqnarray}
\label{eq:xevol}
	\opvec x(t) = \op U(t)^\dag \opvec x_0 \op U(t) = \mat S(t) \opvec x_0\,,
\end{eqnarray}
where all time dependence (or lack thereof) is indicated explicitly, and $\opvec x_0$ is the initial vector of quadratures at ${t=0}$. Correspondingly, the Schr\"odinger evolution of the state, as given by the evolution of the covariance matrix, takes the form
\begin{equation}
	\mat \sigma(t)= \mat S(t)\mat \sigma_0 \mat S(t)^\tp,
\end{equation}
where $\mat \sigma_0$ is the initial state. 

The authors of \cite{Brown} find a differential equation for $\mat S(t)$ that represents the evolution generated by a quadratic Hamiltonian. They first rewrite a general time-dependent, quadratic, Heisenberg-picture Hamiltonian $H$ generating  time translations with respect to the time coordinate~$t$  as
\begin{equation}\label{eq:canon}
	\op H=\opvec x^\tp\mat F(t)\opvec x\,,
\end{equation}
where $\mat F(t)$ is a Hermitian matrix of c-numbers containing any explicit time-dependence of the Hamiltonian. Then they rewrite the Heisenberg equation for the time evolution of the quadratures in terms of the entries of the $\mat F$ matrix:
\begin{equation}
\label{eq:timederivxcomp}
	\frac{d}{dt}\op x_j = i\big[\op H,\op x_j \big] = i\sum_{mn}{F}_{mn}(t)\big[\op x_m \op x_n,\op x_j\big] = \sum_{mn}{F}_{mn}(t) \Bigl( \op x_m \Omega_{jn} + \Omega_{jm} \op x_n \Bigr)\,,
\end{equation}
which can be collected back into vector form as
\begin{equation}
	\frac{d}{dt}\opvec x= \mat \Omega \mat F^{\text{sym}}(t) \opvec x\,,\qquad \mat F^{\text{sym}}=(\mat F+\mat F^\tp)
\end{equation}
Using this into Eq.~\eqref{eq:xevol} we get
\begin{equation}
	\frac{d}{dt}\big[\mat S(t)\big] \opvec x_0=\mat \Omega \mat F^{\text{sym}}(t)\mat S(t)\opvec x _0\,,
\end{equation}
which after some algebra yields  the following first-order, linear, ordinary differential equation for the symplectic matrix:
\begin{equation}
\frac{d}{dt}\mat S(t)=\mat \Omega \mat F^{\text{sym}}(t)\mat S(t).
\end{equation}
Solving this equation with the initial condition $\mat S(0)=\mat I$ such that $\opvec x_0=\mat S(0)\opvec x_0$ is equivalent to solving the standard Hilbert space evolution with the Hamiltonian-unitary formalism.

It is common in the analysis of particle detectors in relativistic settings to write the Hamiltonian in terms of annihilation and creation operators. To quickly show how to compute the coefficients of the $F$ matrix in that case. Let us stack ladder operators on top of each other to form the following column vectors:
\begin{eqnarray}\label{avector}
	\opvec a &\coloneqq (\op a_{d_1},\dots,\op a_{d_M}, \op a_1,\dots \op a_N)^\tp, \nonumber \\
	\opvec a^\dag &\coloneqq (\op a_{d_1}^\dag,\dots,\op a_{d_M}^\dag, \op a_1^\dag,\dots \op a_N^\dag)^\tp. 
\end{eqnarray}
The Hamiltonian from Eq.~\eqref{eq:canon} can be put into the form
\begin{eqnarray} \label{hamil2a}
	\op H = (\opvec a^\dag)^\tp \mat w(t) \opvec a +(\opvec a^\dag)^\tp \mat g(t) \opvec a^{\dag} +\opvec a^\tp \mat g(t)^\herm \opvec a\,,
\end{eqnarray}
where $\mat w(t)$ and $\mat g(t)$ are coefficient matrices, and $\mat M^\herm = \mat M^{*\tp} = \mat M^{\tp*}$ denotes the conjugate transpose of any matrix $\mat M$.

By equating \eqref{hamil2a} to \eqref{eq:canon} and comparing coefficients it is easy to obtain the form of the matrix $\mat F(t)$, which takes the following block form:
\begin{equation}
\mat F(t)=\left(
\begin{array}{cc}
\mat A(t) & \mat X(t)\\
\mat X(t)^\herm& \mat B(t)
\end{array}\right)\,,
\end{equation}
where
\begin{eqnarray}
\mat A(t)&=\frac12\left(\mat w(t)+\mat g(t)+\mat g(t)^\herm\right)\,,\\
\mat B(t)&=\frac12\left(\mat w(t)-\mat g(t)-\mat g(t)^\herm\right)\,,\\
\mat X(t)&=\frac i 2\left(\mat w(t)-\mat g(t)+\mat g(t)^\herm\right)\,.
\end{eqnarray}

Note that we derived these results using global Minkowski time~$t$ because it is the least biased time coordinate when there are multiple detectors involved. But any time coordinate can be used instead---such as some particular detector's proper time---as long as the appropriate transformation is made to the Hamiltonian as detalied in \cite{Brown}.

The model developed in \cite{Brown} has been successfully employed in several important scenarios in quantum field theory and relativistic quantum information, ranging from entanglement harvesting and farming \cite{Brown,farming}, To the analysis of the universality of the Unruh effect in optical cavities \cite{Wilson} and to analyze the causality of the light-matter ninteraction models when UV cutoffs are introduced \cite{Brown,Robort}.

In particular, this model has been proven particularly useful in the anlysis and the predictions  in situations where it is necessary to go beyond perturbation theory. This is the case of a particle detector thermalization under constant acceleration \cite{Wilson}, which allowed the analysis of the universality of the Unruh effect, or the proposal of the settings where field entanglement is harnessed through entanglement farming \cite{farming}. There is work in preparation using this non-perturbative approach to apply the techniques of entanglement-farming to quantum metrology \cite{Brown:2014uj}.

\section{Entanglement in quantum fields}
\label{sec:fieldent}

\subsection{Entanglement depends on tensor-product decomposition}

Generically, the vacuum of a quantum field is entangled~\cite{Unruh1976,Alegbra1,Alegbra2,JuanLocality,Collapse2}. But what does this statement actually mean? We know from results in quantum information theory that whether a state is entangled or not depends on the chosen tensor-product decomposition of the overall Hilbert space~\cite{Zanardi2004}.

In fact, if we choose to decompose the state space of a free scalar field in Minkowski spacetime~\cite{Birrell1982} into plane-wave modes,
\begin{equation}
    \mathcal H_{\text{field}} = \bigotimes_k L^2(\reals)_k\,,
\end{equation}
where $L^2(\reals)_k$ is the countably infinite harmonic-oscillator state space associated with mode~$k$, then the Minkowski vacuum~$\ket 0_\mathrm{M}$ is not entangled at all. In fact, it is nothing but a product state~\cite{Peskin1995}:
\begin{equation}
    \ket{0}_{\textmod{M}} = \bigotimes_k \ket 0_k\,,
\end{equation}
On the other hand, if we decompose the field into a left and right half (corresponding to left and right Rindler wedges; see Section~\ref{Rindbogo}),
\begin{equation}
    \mathcal H_{\text{field}} = \mathcal H_{\text{left}} \otimes \mathcal H_{\text{right}}\,,
\end{equation}
then the Minkowski vacuum is a tensor product of two-mode squeezed~(TMS) states in pairs of \emph{Rindler modes} indexed by~$\omega$~\cite{Unruh1976} (see section \ref{Rindbogo}):
\begin{eqnarray}
    \ket{0}_{\textmod{M}} &=& \bigotimes_\omega \ket {\text{TMS}}_{(\omega,\textmod{I});(\omega,\textmod{II})}\,, %
\end{eqnarray}
whose explicit form is given in Eq.~\eqref{scalarvacuum1}. This decomposition reveals bipartite entanglement across the left-right cut.

We need not stop with a simple bipartition, however. In fact, the picture of quantum fields as a continuum limit of, for example, a bed of coupled oscillators is useful in this case. This is the standard picture when considering quantum fields for condensed matter systems~\cite{Tsvelik:2006vj}, and it provides an intuition for a field as having a state space decomposable into local Hilbert spaces associated to every point in physical space (i.e.,~on some Cauchy surface):
\begin{equation}
    \mathcal H_{\text{field}} = \bigotimes_x L^2(\reals)_x\,
\end{equation}
for bosons. Note that symmetrization is not imposed at the level of the state space but will instead be enforced explicitly by restricting the parameters in the possible representations of field states. (These constructions can be generalised to fermions as long as one is careful with all the subtleties involved in the definition of the tensor-product structure of the fermionic Hilbert space~\cite{Bradlercomment,Edureply,Friis}.)

Recent results for \mbox{(1+1)}-dimensional quantum field theories use exactly this viewpoint: the variational ans\"atze of \emph{continuous matrix-product states}~(cMPS)~\cite{Verstraete:2010bf} and the \emph{continuous multiscale entanglement renormalisation ansatz}~(cMERA)~\cite{Haegeman:2013im} provide a compact, analytic way of representing the state of a quantum field as a path-ordered or time-ordered exponential operator acting on \emph{local} degrees of freedom at every point in space. cMPSs are most useful for representing states of massive fields~\cite{Verstraete:2010bf}:
\begin{equation}
\label{eq:cmpsstate}
    \ket{\text{field state}} = \tr_{\text{aux}} \left\{ \mathcal P \exp \left[ \int dx\, \Bigl( Q(x) \otimes 1 + R(x) \otimes \psi^\dag(x) \Bigr) \right] \right\} \ket \Omega\,,
\end{equation}
where $\ket\Omega$ is the tensor product of the \emph{local ground states} of all local field degree of freedom (each with annihilation operator~$\psi(x)$) at every point%
, $Q(x)$ and $R(x)$ are spatially varying ${D\times D}$ matrices that act on a (fictitious) auxiliary system, $\mathcal P \exp$ denotes the path-ordered exponential, and $\tr_{\text{aux}}$ denotes tracing out the auxiliary system (i.e.,~taking the matrix trace over the $D$-dimensional space on which $Q$ and $R$ act). For \mbox{(1+1)}-dimensional conformal fields, cMERA is more appropriate than cMPS because it has scale invariance explicitly built in~\cite{Haegeman:2013im}:
\begin{equation}
\label{eq:cmerastate}
    \ket{\text{field state}} = \mathcal T \exp \left[ -i \int ds\, \Bigl( K(s) + L \Bigr) \right] \ket \Omega\,,
\end{equation}
where $K(s)$ is a local entangling interaction, $L$ is a scale change operation (dilation), and $\mathcal T \exp$ is the time-ordered exponential. Recall that~$\ket\Omega$ is not the (global) vacuum of the full field; it is merely the tensor product of the ground state of each local degree of freedom. The vacuum of a quantum field therefore also admits a description in terms of Eq.~\eqref{eq:cmpsstate} or Eq.~\eqref{eq:cmerastate}, as appropriate to the type of field. When viewed in this way, the interpretation of the vacuum as an multipartite entangled state over the tensor product of many local degrees of freedom naturally follows.

The purpose of this subsection is to remind the reader explicitly that entanglement is only defined with respect to a (freely chosen) tensor-product decomposition of the full state space. Statements about ``entanglement in the field'' must therefore be understood within this context. Ref.~\cite{Zanardi2004} provides guidelines for choosing an appropriate tensor-product decomposition based on which laboratory observations are ``easy'' to do. This is usually---but not always---related to which ones can be implemented locally. On the other hand, sometimes mathematical considerations might lead one to choose a different decomposition whose physical relevance is not immediately apparent (e.g.,~Rindler modes or Unruh modes~\cite{Unruh1976}). In that case, one should be careful when making physical claims without first translating back to what measurements could be made on such a system.

As a particular example of this, the presence or absence of entanglement in a quantum field does not necessarily translate directly to what local observers would detect. The question of how to extract---or \emph{harvest}~\cite{MartinMartinez2012}---entanglement from a quantum field using local detectors will be discussed in Sec.~\ref{sec:entharvest}. In the rest of this section, we will examine a few examples of entanglement within quantum fields. In doing so, one should keep in mind the aforementioned caveats of physical interpretation and applicability.

\subsection{Bosonic versus fermionic field entanglement in cosmology}
\label{subsec:bosfermfieldent}

Even when looking at the entanglement contained in quantum fields in the simple flat spacetime case, there are strong differences between the behaviour of fermionic and bosonic entanglement. These differences were already pointed out in  the pioneering works on field  entanglement in non-inertial frames, where it was shown that fermionic and bosonic entanglement behaves in a completely different way \cite{Alsingtelep,Alicefalls,Fuentes2010}. These differences were better understood and related to the field statistics through later studies \cite{Edu4,beyond,Edu3,Edueks,Arbitrary,Friis2011}. In \cite{Miguel1}, the authors  pointed out that there were intrinsic ambiguities related to the formalism employed in the original works studying fermionic entanglement. This comes about because all those works effected a map between fermionic field states into a bosonic Hilbert space endowed with a convenient tensor product structure.  Reference~\cite{Miguel1}  shows the problem and proposes an effective  solution too, but it  raised serious concerns about the interpretation of the previous results. Further work helped build  the correct intuition \cite{Bradlercomment,Edureply,Friis}. These works allowed for a better understanding of the behaviour of correlations in non-inertial frames. This improved understanding lead to several unanticipated results. For example, (1)~the fact that, contrary to believe, it is not necessarily true that higher accelerations imply degradation of field entanglement \cite{MigC}; and (2)~the obtention of new and more consistent insight into the nature of fermionic field entanglement in extreme limits of high acceleration \cite{Miguel3}.

The differences between different field statistics are also present in the phenomenon of particle creation due to the expansion of the universe in scenarios where conformal symmetry is broken. It has been previously analyzed and compared \cite{MartinMartinez2012} the difference in the quantum correlations present in the particle creation due to the the expansion of spacetime in the case of massive bosonic \cite{caball} and fermionic~\cite{fermxpanding} fields. These results were also reviewed in \cite{MartinMartinez2012}, so in the following, we provide a short summary.

 In the minimal coupling scenario, the covariant form of the Dirac equation is
\begin{equation}
(\ii e^\mu_a\tilde\gamma^aD_\mu+m)\psi=(\ii\gamma^{\mu}D_\mu+m)\psi=0\,.
\end{equation}
We can denote particle and antiparticle solutions of momentum~$k$ in the asymptotic past, respectively, as $u^+_{\mathrm{in}}(k)$ and $u^-_{\mathrm{in}}(k)$. In the same fashion, the particle and antiparticle solutions in the asymptotic future will be denoted as $u^+_{\mathrm{out}}(k)$ and $u^-_{\mathrm{out}}(k)$.

In  a FLRW spacetime, the vierbein takes the form $e^{a\mu}=\sqrt{C(\eta)}\eta^{a\mu}$, where $C(\eta)=[a(\eta)]^2$ is the conformal factor. As discussed in previous sections, given the  conformal symmetry of the minimal coupling in 1+1 dimensions (up to mass terms), we can relate the Dirac equation in the FLRW scenario to its flat-spacetime form. Using this we find, as in the scalar case, that the Bogoliubov transformation between modes in the asymptotic past and future does not mix solutions of different momentum:
\begin{equation}
u_{\mathrm{in}}^\pm(k)= \alpha_{k}^{\pm}u_{\mathrm{out}}^\pm(k)+\beta_{k}^{\pm}u_{\mathrm{out}}^{\mp*}(k)\,.
\end{equation}

If we now assume a specific form for the conformal factor, the Bogoliubov coefficients can be computed.  As discussed in \cite{MartinMartinez2012}, the same form that allowed us to obtain analytical solutions in the bosonic case will not do so in the fermionic case. Instead we would need the vierbein field (and not the metric as in the bosonic case) to be proportional to $1+\epsilon\tanh(\rho\eta)$. This means that if we choose a form for the conformal factor given by
\begin{equation}
C(\eta)=[1+\epsilon\tanh(\rho\eta)]^2\,,
\end{equation}
Compating this expression with Eq.~\eqref{factorbos}, we see that the spacetime acquires exactly the same interesting properties as in the scalar scenario. It describes a process of asymptotic flatness followed by a smooth transition to fast expansion and later followed by a smooth transition to asymptotic flatness again at a larger scale factor. The asymptotic flatness at both ends allows us to define  `in' and `out' particle states. The similarity between this spin-$\frac 1 2$ case and the scalar case allows us to directly compare the qualitative results obtained in each.
 
Following a process completely analogous to what lead us to Eq.~\eqref{uno1} and Eq.~\eqref{uno2}, we obtain the form of the Bogoliubov coefficients, and from them we obtain the form of the `out' annihilation operators in terms of `in' annihilation and creation operators, following the process detailed in~\cite{MartinMartinez2012,fermxpanding}:
\begin{eqnarray}
\label{opF1}a_{{\mathrm{out}},\bm k,\sigma}&=
\frac{K_{{\mathrm{in}}}}{K_{{\mathrm{out}}}}\Bigg(\alpha_{k}^{-}a_{{\mathrm{in}},\bm k,\sigma}+\beta_{k}^{-\ast}\sum_{\sigma'}
X_{\sigma'\sigma}(-{\bm k})b^{\dag}_{{\mathrm{in}},-\bm k,\sigma'}\Bigg)\,, \\*
 \label{opF2}b_{{\mathrm{out}},\bm k,\sigma}&=
\frac{K_{{\mathrm{in}}}}{K_{{\mathrm{out}}}}\Bigg(\alpha_{k}^{-}b_{{\mathrm{in}},\bm k,\sigma}-\beta_{k}^{-\ast}\sum_{\sigma'}
X_{\sigma\sigma'}({\bm k})a^{\dag}_{{\mathrm{in}},-\bm k,\sigma'}\Bigg)\,,
\end{eqnarray}
where $\sigma=\pm$. The polarisation tensor then takes the form
 \begin{equation}
X_{\sigma\sigma'}({\bm k})=-2\mu_{\mathrm{out}}^2K^2_{{\mathrm{out}}}\bar{U}_{{\mathrm{out}}}(-{\bm k},\sigma')V(0,\sigma)\,,
\end{equation}
where
 \begin{equation}
K_{{\mathrm{in}}/{\mathrm{out}}}=\frac{1}{|\bm k|}\sqrt{\frac{\omega_{{\mathrm{in}}/{\mathrm{out}}}(\bm k)-\mu_{\mathrm{in}/\mathrm{out}}}{\mu_{\mathrm{in}/\mathrm{out}}}}\,,
\end{equation}
$\bar{U}_{{\mathrm{out}}}({\bm k},\sigma)$ is a spinor solution of the Dirac equation in the curved background, and $V(\bm k,\sigma)$ is a flat-spacetime antiparticle spinor~\cite{dun1}. 
 
If we operate as we did with the scalar case and consider 1+1 dimensions, a and regard the Dirac field a Grasmann scalar field~\cite{Arbitrary}. In this case, Eq.~\eqref{opF1} and Eq.~\eqref{opF2} are greatly simplified~\cite{fermxpanding}. With the same technique used to obtain Eq.~\eqref{dynoresult}, we obtain the form of the vacuum state in the asymptotic past in terms of particle and antiparticle modes in the asymptotic future demanding that the operators $a_{{\mathrm{in}},\bm k,\sigma}$ and $b_{{\mathrm{in}},\bm k,\sigma}$ annihilate the `in' vacuum in the `out' basis. This yields the following expression for the `in' vacuum state in terms of `out' modes:
\begin{equation}
|0\rangle_{\mathrm{in}}=\bigotimes_{k}\frac{1}{\sqrt{1+|\theta_{\mathrm{F}}|^{2}}}\Big(|0\rangle_{{\mathrm{out}}}-\theta_{\mathrm{F}}\,b^\dagger_{\mathrm{out},k}a^\dagger_{\mathrm{out},-k}|0\rangle_{{\mathrm{out}}} \Big),\qquad
\theta_{\mathrm{F}}=\frac{\beta_k^{-*}}{\alpha_k^{-*}} \frac{\mu_{\mathrm{out}}}{|k|}\left(1-\frac{\omega_{\mathrm{out}}}{\mu_{\mathrm{out}}}\right)\,.
\end{equation}
The entanglement entropy of this state was obtained in Ref.~\cite{fermxpanding} as
\begin{equation}
S_{\mathrm{F}}= \log\left(\frac{1+|\theta_{\mathrm{F}}|^{2}}{|\theta_{\mathrm{F}}|^{
\frac{2|\theta_{\mathrm{F}}|^{2}}{|\theta_{\mathrm{F}} |^{2}+1}}}\right)\,,
\label{SF}
\end{equation}

Both for the bosonic and the fermionic case the entropy is zero when the mass of the field vanishes, this being the result of an equivalence between minimal and conformal coupling in 1+1 dimensions. In that case the `in' vacuum would be simply the conformal vacuum, which, as discussed above, remains invariant under time evolution.

\begin{figure*}[t]
\begin{center}
\includegraphics[width=.45\columnwidth]{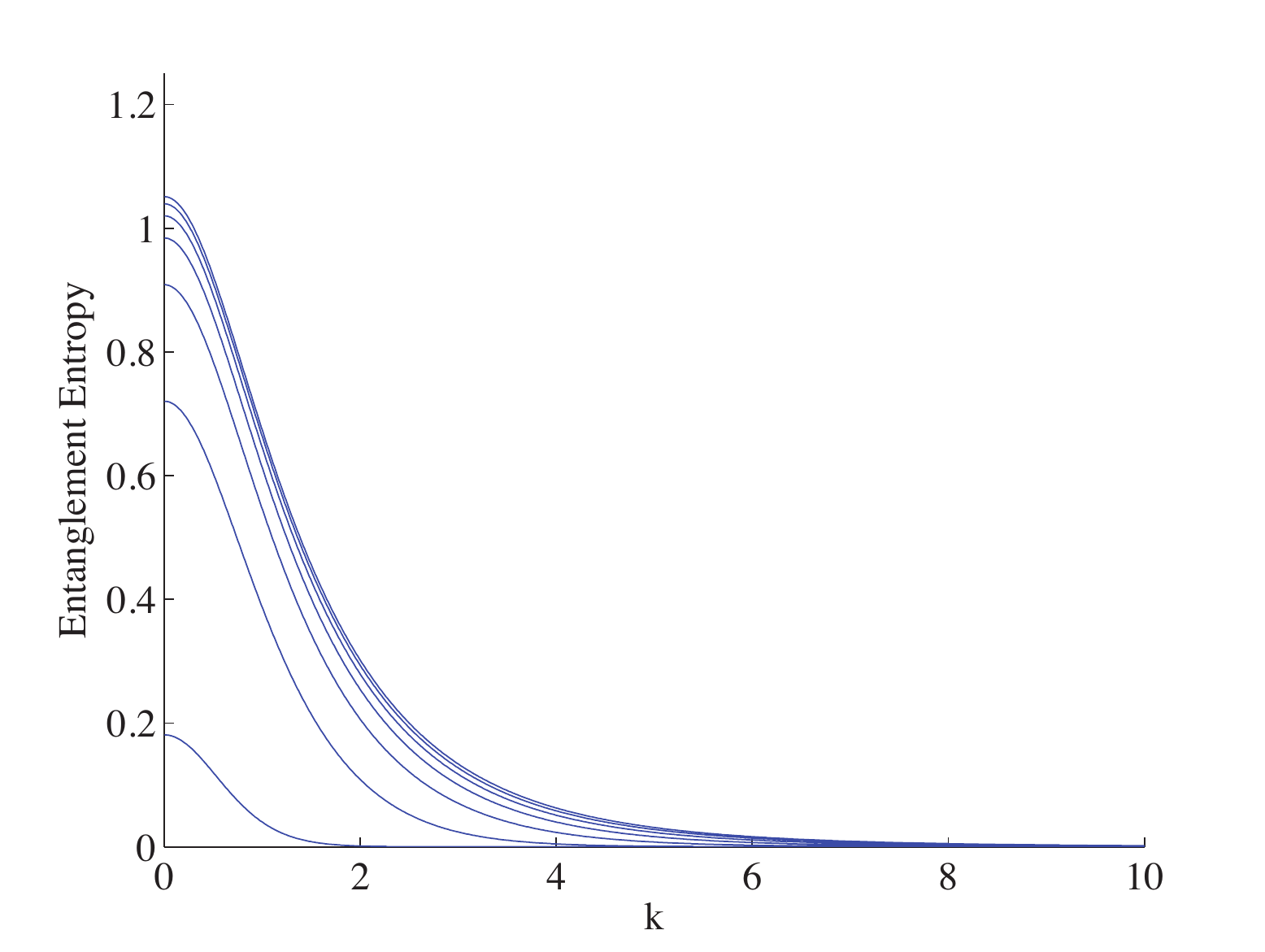}
\includegraphics[width=.45\columnwidth]{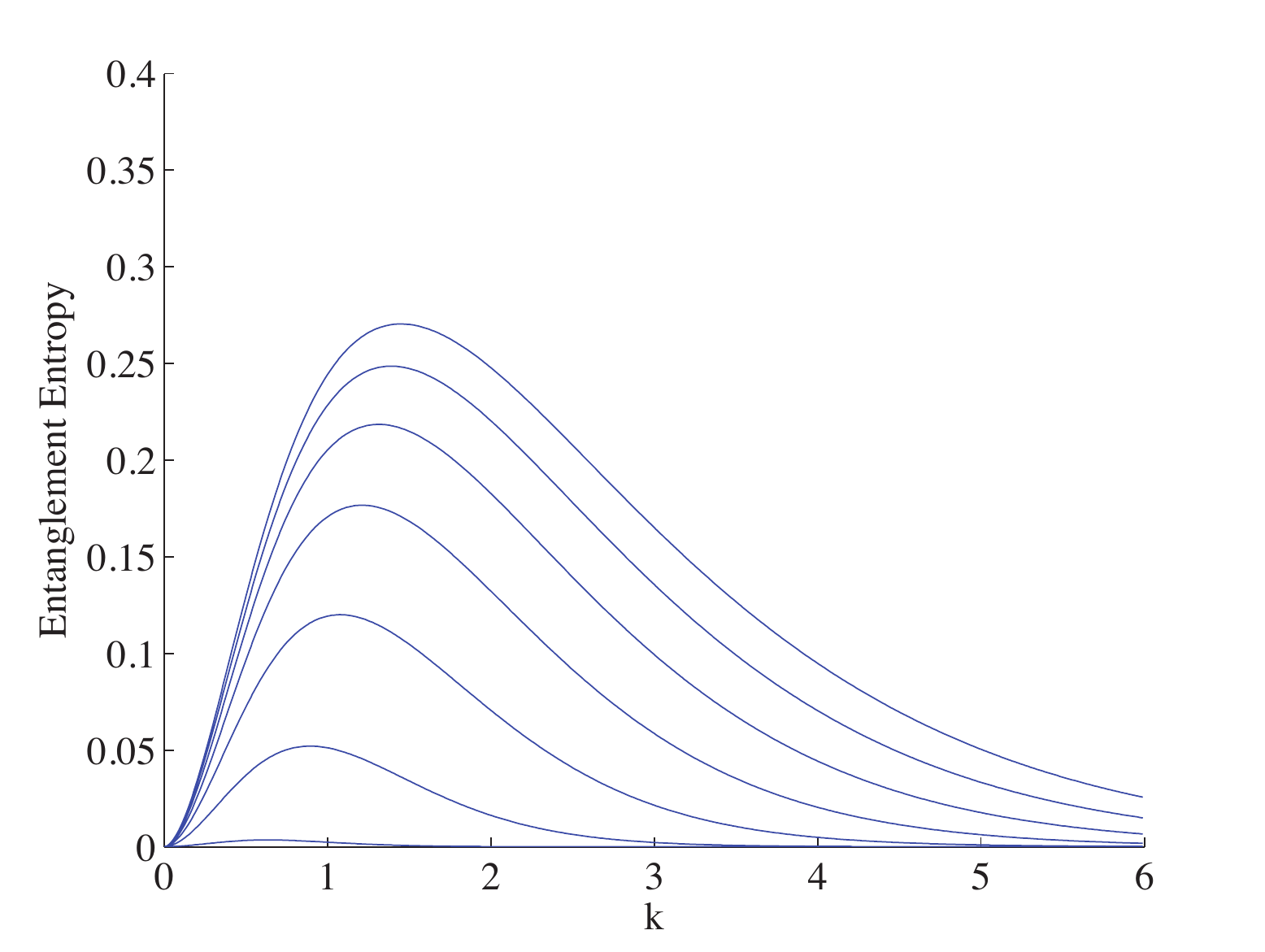}
\end{center}
\caption{{[Based on figures in Ref.~\cite{fermxpanding}]} Bosonic field (\textbf{left}) and fermionic field (\textbf{right}): Entropy of entanglement for a fixed mass $m=1$ as a function of $|\bm k|$ for different rapidities $\rho=1,\dots,40$, with a fixed value of $\epsilon=1$.}
\label{differences}
\end{figure*}

When  conformal symmetry is broken, both scalar and Dirac field modes become entangled due to the expansion of the universe. In both cases, the amount of entanglement generated in the field due to the expansion codifies information about the underlying spacetime. However, as analysed in Refs.~\cite{MartinMartinez2012,fermxpanding}, there are fundamental differences in the spectral behaviour of that entanglement. In Fig.~\ref{differences}, it is shown how the entanglement depends on the momentum of the field mode for different values of the parameter~$\rho$. For the Dirac case, entanglement peaks at a certain momentum, while for the scalar field, entanglement monotonically decreases with momentum and has its maximum at $|k|=0$. In contrast to the scalar field, for the Dirac field, there is a privileged value of $|k|$ for which the expansion of the spacetime generates a large amount of entanglement. Modes of that characteristic frequency are far more prone to entanglement than any others.

In Ref.~\cite{fermxpanding}, an attempt to interpret this phenomenon is given. It is based on the fact that the optimal value of $|\bm k|$ can be associated with a characteristic wavelength (proportional to $|\bm k|^{-1}$) that is increasingly correlated with a characteristic length of the universe. Intuitively, fermion modes with higher characteristic lengths are less sensitive to the underlying spacetime because the exclusion principle impedes the excitation of very long-wavelength modes---i.e.,~those whose $|\bm k|\rightarrow0$. For bosons, where this constraint does not exist, the entanglement generation is higher when $|\bm k|\rightarrow0$. This can be intuitively explained by the fact that modes of smaller $|\bm k|$ are more easily excited as the spacetime expands since it is energetically much cheaper to excite smaller $|\bm k|$ modes.

This natural emergence of a privileged mode in the fermion field is very sensitive to the expansion parameters and thus more efficiently encodes information about the underlying spacetime than in the scalar case as discussed In Refs.~\cite{fermxpanding,MartinMartinez2012}. While the expansion of the universe without conformal symmetry has been proven to generate entanglement regardless of the boson or fermion nature of the quantum field, we have seen that Dirac fields seem to codify more information about the underlying expansion, revealing that field statistics can play a key role in the way in which the expansion of the universe generates entanglement in quantum fields.

\section{The echo of the early universe in vacuum fluctuations}
\label{sec:entinflation}

It has been recently proposed that the vacuum fluctuations of a quantum field, observed by comoving observers, may contain information about the very early stages of the Universe, and that this information could be readable even nowadays. In~\cite{echo} this question has been analyzed employing a toy model inspired by curved-spacetime QFT, quantum optics and relativistic quantum information~\cite{Wavepackets}. Namely,~\cite{echo} estimates the signatures of the early-universe spacetime background on a quantum system which was coupled to a scalar field during the early stages of the universe and whose interaction has remained switched on until our era. 

In summary, in this work the authors estimate the impact that deviations from the predictions of general relativity may have on interactions that break conformal symmetry during early times (where the energy density was comparable to the Plank scale).  They make use of an Unruh-DeWitt detector interaction model and evaluate the sensitivity and the memory of the detector's vacuum response to Planck-scale physics under the most conservative possible hypothesis. With this in mind, they analyze the response of such an idealised particle detector which has remained coupled to matter fields from the early stages of the Universe until today. The authors pose the question whether this detector would conserve any information from the time when it witnessed the very early Universe dynamics, encoded in its vacuum response.

Intuitively, one would think that any effect imprinted on the response of the detector in the early Universe would have been most likely washed out after the overwhelming amount of time from the Planck scale to our era. Hence, it may seem that there is little hope in finding any trace of early Universe physics in the vacuum response of the particle detector today. Consequently, the idea of looking for signatures of quantum gravitational effects on a particle detector  that  was coupled to matter during some interesting period of time close to the Plank scale---and then remained coupled for the entire life of the  Universe---may  sound, at first sight, far fetched. The authors of~\cite{echo} challenge this statement.

The authors compare two different scenarios, one in which the universe dynamics is dictated by classical General Relativity (GR), with another where they adopt the effective dynamics predicted by Loop Quantum Cosmology (LQC)~\cite{Bojowald:2008zzb,Ashtekar:2011ni,Banerjee:2011qu,Agullones}. Both dynamics disagree only when matter-energy densities were comparable to the Planck scale.

Note that the use of Loop Quantum Cosmology in this work is of little relevance and constitutes just a convenient means to assess the impact of a different spacetime dynamics from that predicted by GR during the early stages of the universe. Indeed, the conclusion of the paper is independent of the validity or not of LQC as a reasonable model of quantum cosmology. Even if the correct theory of gravitation does predict that there is no such thing as a spacetime geometry near the Plank scale, one would still expect that there have to be intermediate energy scales where a semi-classical theory, stemming from effective quantum corrections to Friedmann equations, would predict, for a short time, an effective spacetime geometry that deviates from GR though the effective quantum corrections, much in the same manner as post-Newtonian gravity approximates general relativity in intermediate energy scales. 

\begin{figure}[ht]
\begin{center}
\includegraphics[width=0.65\textwidth]{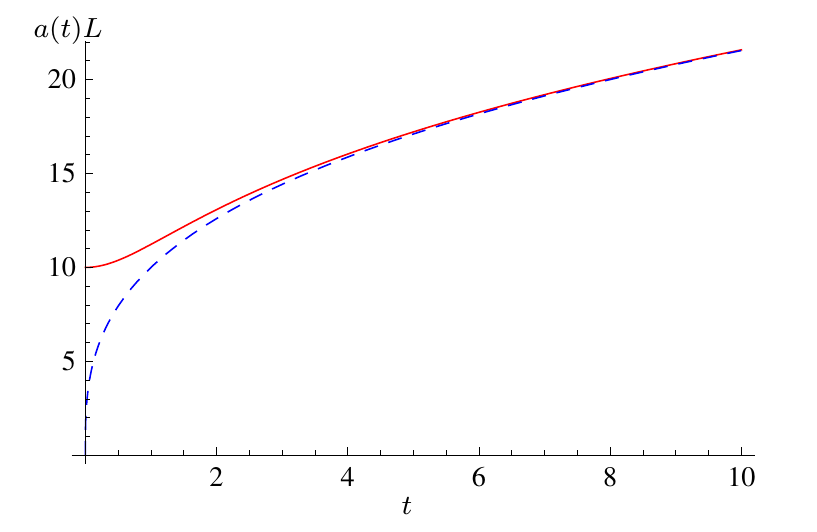}
\caption{Scale factor as a function of the proper time for $l=1$, 
$\pi_\varphi=1000$. The dashed blue curve represents the classical scale factor 
$a_{\mathrm{GR}}(t)L$ and the solid red curve corresponds to the LQC effective scale factor 
$a_\mathrm{LQC}(t)L$. All quantities are expressed in Planck units, i.e. 
$\lplanck=1$.}
\label{fig:scalefactors}
\end{center}
\end{figure}

As stated above, the authors compare classical General Relativity  with effective Loop Quantum Cosmology.  For that, they consider a spatially flat, homogeneous and isotropic Universe, $ ds^2=~a(t)[-d\eta(t)^2+d \vec{x}^2]$, with  a massless scalar $\varphi$ as matter source. The scale factor for each dynamics is~\cite{echo}
 \begin{eqnarray}\label{class}
a_{GR}(t)=\frac{( \lplanck^2\pi_\varphi^2)^{1/6}}{L}t^{1/3},
\quad 
a_{LQC}(t)=\frac{l}{L}\left(\frac{\pi_\varphi^2}{{\lplanck^2}}\right)^{1/6}\left[1+\left(\frac{\lplanck^2}{l^3}\right)^2t^2\right]^{1/6},
\end{eqnarray}
and the conformal time $\eta$ in terms of the comoving time $t$ is
\begin{eqnarray}\label{etaclass}
\eta_{\mathrm{GR}}(t)=\frac{3L\,t^{2/3}}{2(\lplanck^2\pi_\varphi^2)^{1/6}},
\quad
\eta_{LQC}(t)=\frac{L}{l}\left(\frac{\lplanck^2}{\pi_\varphi^2}\right)^{1/6}t \cdot {_2}F_1\left[\frac1{6},\frac1{2},\frac{3}{2},-\left(\frac{\lplanck^2}{l^3}t\right)^2\right].
\end{eqnarray}
Here, $\lplanck=\sqrt{12\pi G}$ is the Planck length,
$\pi_\varphi$ is the momentum conjugate to $\varphi$ and is a constant of motion, $l$ is a quantization parameter (in LQC the volume has a discrete spectrum  equally spaced by $2l^3$ units)
\cite{Ashtekar:2011ni,Banerjee:2011qu}, and ${}_2F_1$ is an ordinary hypergeometric function.  They consider a three-torus spatial topology, with coordinates in the interval $[0,L]$. For compactification scales larger than the observable Universe, this flat topology is compatible with  observations~\cite{Mukhanov:2005sc}. The classical GR solution,  $a_{\mathrm{GR}}(t)=~\lim_{l\rightarrow 0} a_{\mathrm{LQC}}(t)$, displays a big-bang singularity at 
 $t=0$. In contrast, $a_{\mathrm{LQC}}(t)$ never vanishes (see Fig.~\ref{fig:scalefactors}). LQC   replaces this singularity  by a big-bounce, i.e.,  the Universe shrinks for $t<0$, bounces at $t=0$, and expands for $t>0$. In the limit $t\gg{l^3}/\lplanck^2$, $\eta_{\mathrm{LQC}}(t)\rightarrow \eta_{\mathrm{GR}}(t)+\beta$ with
 
\[\beta=\frac{l^2L\sqrt{\pi}\,\Gamma\left(-\frac1{3}\right)}{(\pi_\varphi^2 \lplanck^{10})^{1/6}2\Gamma\left(\frac1{6}\right)}\]

On top of this background, the authors  consider a conformal massless scalar field $\phi$  filling the Universe.  The choice of both the coupling to curvature and the initial state of the field  respond to a conservative hypothesis: As discussed in Sec \ref{qubitUDW}, if the initial state of the field is different from the conformal vacuum, there is particle creation due to the expansion of the universe. This will be dependent on the early-universe dynamics and contribute to increase the signature on the detector. If we want to be conservative with respect to the vacuum noise signature on the detector, we have to choose as initial state the one that minimises particle creation. The most conservative hypothesis is thus to consider the state of the field to be the conformal vacuum, which remains invariant as the Universe expands~\cite{cosmoq}. 
The field operator is $\hat{\phi} =\sum_{\vec{n}} (a u_{\vec{n}} +a^\dagger u^*_{\vec{n}})$ with $a$ ($a^\dagger$) the corresponding annihilation (creation) operators. This Fock  quantization  is the only  one  (up to unitary equivalence) with a vacuum invariant under the spatial isometries and unitary quantum dynamics~\cite{Gomar:2012xn,Gomar:2012pp}. 
 In the above expression,  note that $\bm n\neq \bm 0$. As  discussed in~\cite{Gomar:2012xn,Gomar:2012pp},  in~\cite{echo}  the zero mode dynamics is ignored. This is done assuming that the state of the zero mode can be kept under control. this is a subtle point  deeply analyzed in~\cite{JormaEdu}, where it is discussed under what circumstances this is a reasonable hypothesis. In~\cite{echo}, the authors assume that the detector is not coupled to this mode, restricting the sums to $\bm n\neq \bm 0$. The unitary evolution and the uniqueness of the representation do not depend on the removal of a finite number of degrees of freedom. Furthermore, one can see that the coupling of the detector to the zero mode $\vec n=\bm 0$ of the field is not relevant to the effects reported in this article if considered in a `safe' state as described in~\cite{JormaEdu}. Something that can be readily seen by, for instance, introducing a small field mass or an IR regularization, or  by coupling the detector to the derivative of the field instead of the field itself~\cite{IRdet,Benitoooooo} (a model that is free from IR divergencies and for which the zero mode influence can be more easily minimised~\cite{JormaEdu}).

Now, the proper time of comoving observers (who see an isotropic expansion)  does not coincide with the conformal time. Hence,  comoving detectors actually detect particles even in the conformal vacuum~\cite{cosmoq}.  This is the well-known Gibbons-Hawking effect~\cite{GibHawking}, which we analyze to identify signatures of quantum gravitational effects on the particle detector (See section \ref{qubitUDW}).

The authors consider an Unruh-DeWitt detector stationary in the comoving frame, $\vec{x}(t)=\vec x_0$, and initially in its ground state. Recall that the Hamiltonian of the coupled system  in the interaction picture is $\hat{H}_I(t)=~\lambda\;\chi(t)(\sigma^+ e^{i\Omega t}+\sigma^- e^{-i\Omega t})\hat{\phi}[\vec{x}(t),\eta(t)]$, where $\lambda$ is the coupling strength, $\chi(t)$ is the detector's switching function, $[\vec{x}(t),\eta(t)]$ represents the detector's worldline, $\Omega$ is its energy gap, and $\sigma^\pm$ are $SU(2)$ ladder operators. Assuming a small enough $\lambda$, we can  compute  the transition probabilities for the detector switched on at $T_0$ to be excited at time $T$ within perturbation theory:
\begin{eqnarray}
\label{prob}
P_\mathrm{e}(T_0,T)& =\lambda^2 {\sum_{\vec{n}}}|I_{\vec{n}}(T_0,T)|^2 +\mathcal{O}(\lambda^4),\\
\label{none}
I_{\vec{n}}(T_0,T)&=\int_{T_0}^T \mathrm{d}t 
\frac{\chi(t) }{a(t)\sqrt{2\omega_{\vec n}L^3}}e^{-\frac{ 2\pi\ii 
\bm n\cdot \bm x_0}{L} }e^{\ii[\Omega t + \omega_{\vec{n}} 
\eta(t)]}.\end{eqnarray}
Here, $\omega_{\vec{n}}=\frac{2\pi}{L}\left|{\vec{n}}\right|$, and \mbox{$\vec{n}=(n_x,n_y,n_z)\in\mathbb{Z}^3$.}

To see how sensitive the response of the detector is to the early universe physics, we can compare the probabilities  for the detector to get excited when the Universe evolves under LQC dynamics (with quantization length $l$) and GR (the limit $l\rightarrow 0$ )---i.e., under the dynamics $P^{\mathrm{LQC}}_\mathrm{e}(T_0,T)$ and $P^{\mathrm{GR}}_\mathrm{e}(T_0,T)$, respectively. We are going to divide the integrals in two intervals. Consider that 
$T_\mathrm{m}\in (T_0,T)$ is a short time sufficiently large for $\eta_\mathrm{LQC}(T_\mathrm{m})\approx\eta_{\mathrm{GR}}(T_\mathrm{m})+\beta$,  typically  few times $l^3/\lplanck^2$. Then we split the GR and LQC integrals (both of the form \eqref{none}) into two intervals:  a tiny interval $t\in [T_0,T_\mathrm{m}]$ (comparable to the Planck sale), where LQC and GR  appreciably predict different dynamics, and $t\in [T_\mathrm{m},T]$, where both dynamics are essentially the same. This gives
\begin{eqnarray}
P^{\mathrm{GR}}_\mathrm{e}(T_0,T)&=\lambda^2{\sum_{\bm n}}\bigg|I_{\bm 
n}^\mathrm{GR}(T_0,T_\mathrm{m})+I_{\bm 
n}^\mathrm{GR}(T_\mathrm{m},T)\bigg|^2+\mathcal{O}(\lambda^4),\\
P^{\mathrm{LQC}}_\mathrm{e}(T_0,T)&=\lambda^2{\sum_{\bm n}}\bigg|I_{\bm 
n}^\mathrm{LQC}(T_0,T_\mathrm{m})+e^{\ii\beta\omega_{\bm n}}I_{\bm 
n}^\mathrm{GR}(T_\mathrm{m},T)\bigg|^2+\mathcal{O}(\lambda^4).
\end{eqnarray}
The first shocking observation is that the difference of the detector's particle counting in both scenarios, which we call $\Delta P_\mathrm{e} (T_0,T)$, will be considerable, even for 
$T\gg l^3/\lplanck^2$---that is, if we look at the detector nowadays. If we compute the difference in the particle counting of the detector in both scenarios we get 
\begin{eqnarray}
\Delta P_\mathrm{e}(T_0,T)&\equiv P^{\mathrm{LQC}}_\mathrm{e}(T_0,T)-P^{\mathrm{GR}}_\mathrm{e}(T_0,T)=\lambda^2{\sum_{\bm n}}\bigg[\left|I_{\bm 
n}^\mathrm{LQC}(T_0,T_\mathrm{m})\right|^2-\left|I_{\bm n}^\mathrm{GR}(T_0,T_\mathrm{m})\right|^2 
\nonumber\\
& +2 \mathrm{Re}\Big({I_{\bm n}^\mathrm{GR}}^*(T_\mathrm{m},T)\Big[e^{-\ii\beta\omega_{\bm n}}I_{\bm 
n}^\mathrm{LQC}(T_0,T_\mathrm{m})-I_{\bm n}^\mathrm{GR}(T_0,T_\mathrm{m})\Big]\Big)\bigg]+\mathcal{O}(\lambda^4).
\end{eqnarray}
 This tells us that even if $\left|I_{\bm 
n}^\mathrm{GR}(T_0,T_\mathrm{m})\right|^2- \left|I_{\bm 
n}^\mathrm{LQC}(T_0,T_\mathrm{m})\right|\ll 1 $, the difference between the two dynamics during  the early universe stage gets multiplied by the long time integral ${I_{\bm n}^\mathrm{GR}}^*(T_\mathrm{m},T)$.

Now one can study how sensitive  the response of the detector is to the LQC quantum parameter $l$  that characterises the size of the quantum of volume ($l^3$). With this aim the authors use as estimator the mean relative difference between probabilities of excitation averaged over a long interval in the late time regime  $\Delta T=T-T_{\mathrm{late}}$, with $\Delta T,\;T_\mathrm{late}\gg l^3/\lplanck^2$:
\begin{eqnarray}\label{estimator} 
E=\left\langle\frac{\left\langle  \Delta P_\mathrm{e}(T_0,T) 
\right\rangle_{\raro{T}}}{\left\langle  P^\mathrm{GR}_\mathrm{e}(T_0,T)
\right\rangle_{\raro{T}}}\right\rangle_{\Delta T}.\end{eqnarray}
This estimator tells us the difference in 
magnitude between the number of clicks of the detector in the GR and LQC backgrounds. 
The internal averages  in \eqref{estimator}  are given by
\begin{equation}\label{ave}
\left\langle   P_\mathrm{e}(T_0,T) \right\rangle_{\raro{T}}=\frac{1}{\raro{T} }\int^T_{T-\raro{T}} P_\mathrm{e}(T_0,T')\,\mathrm{d}T',
\end{equation}
where $\raro{T}\gg l^3/\lplanck^2$ is the time resolution with which we can probe the 
detector.  This is so because it is not possible to resolve times as small as $l^3/\lplanck^2$ 
(roughly, the  Planck scale), so any observation  made on particle 
detectors today will necessarily be averaged over many Planck times.  Moreover, in order to remove any 
possible spurious effects of the big differences in the  
scales of the problem, the authors consider a particle detector with a subplanckian energy gap 
$\Omega\ll\lplanck^2/l^3$.   One may legitimately wonder if these practical considerations may destroy the early Universe signal. Indeed, these constraints 
 partially erase the observable difference between the response of the 
detector in the two scenarios.

\begin{figure}
\begin{center}
\includegraphics[width=0.60\textwidth]{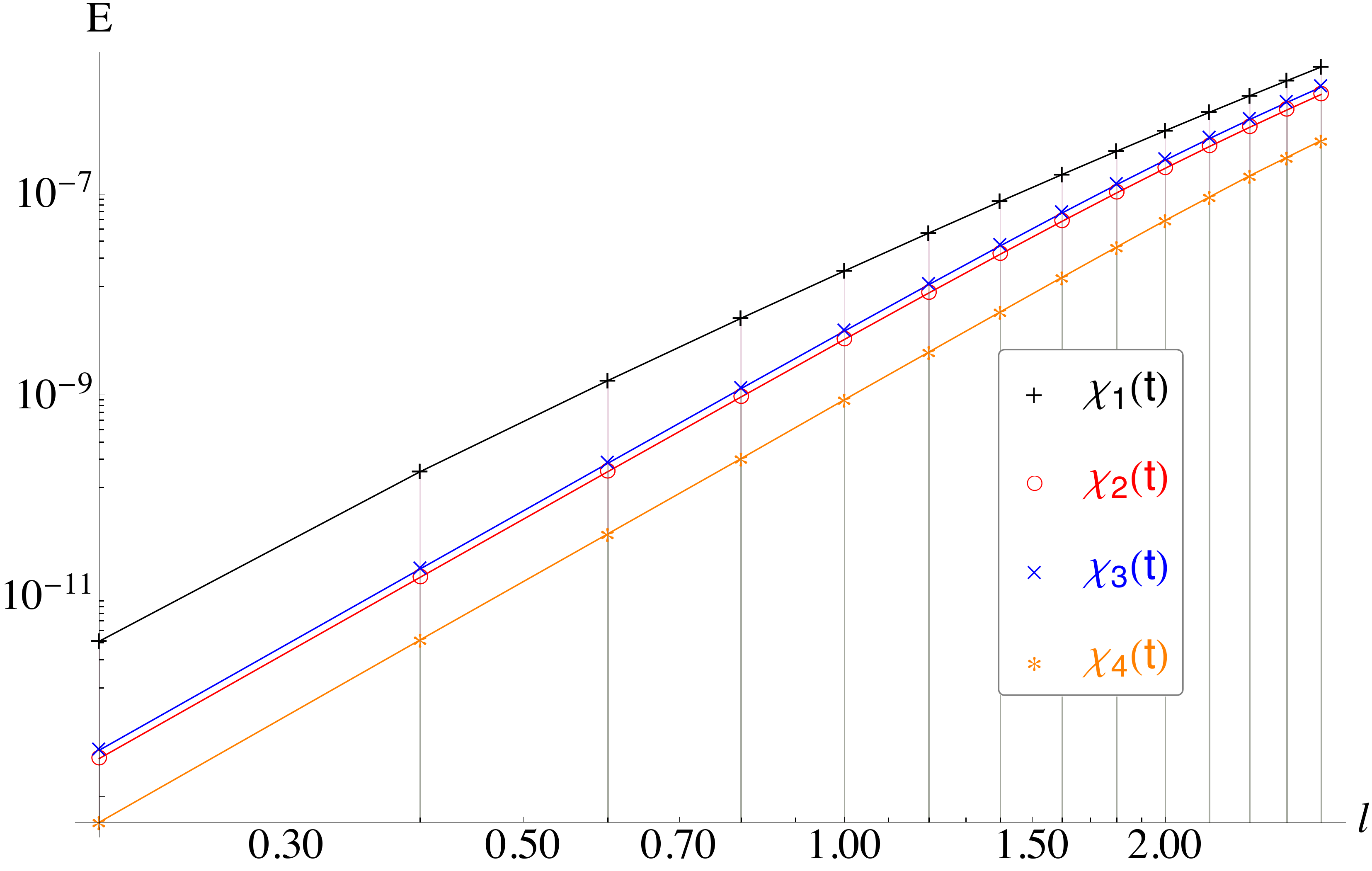}
\caption{Logarithmic plot of the relative difference of the averaged probabilities, E, as a function of the parameter $l$, for the 4 different switching functions \eqref{redios}, \eqref{redios1}, \eqref{redios3}, \eqref{redios4}; and for $\Omega\ll \lplanck^2/l^3$ and $\pi_\varphi=1000$.  The detector is switched on at $T_0=0.01$ (some early time after the bounce). The variation of the response of the detector (the intensity of Gibbons-Hawking type quantum fluctuations) grows exponentially with l.}
\label{fig:tomajeroma}
\end{center}
\end{figure}

\begin{figure*}[t]
\begin{center}
\includegraphics[width=.30\textwidth]{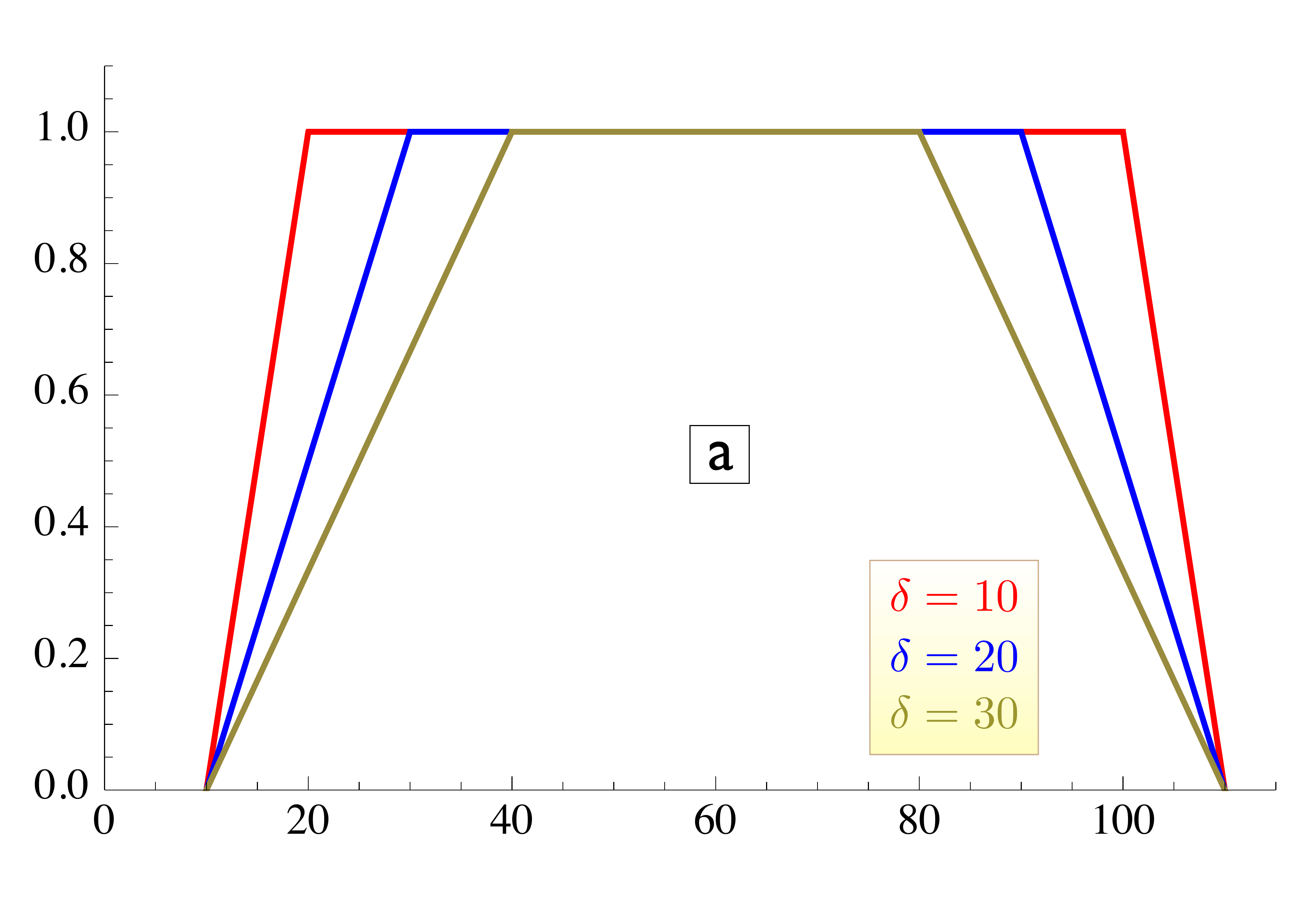}\,\includegraphics[width=.30\textwidth]{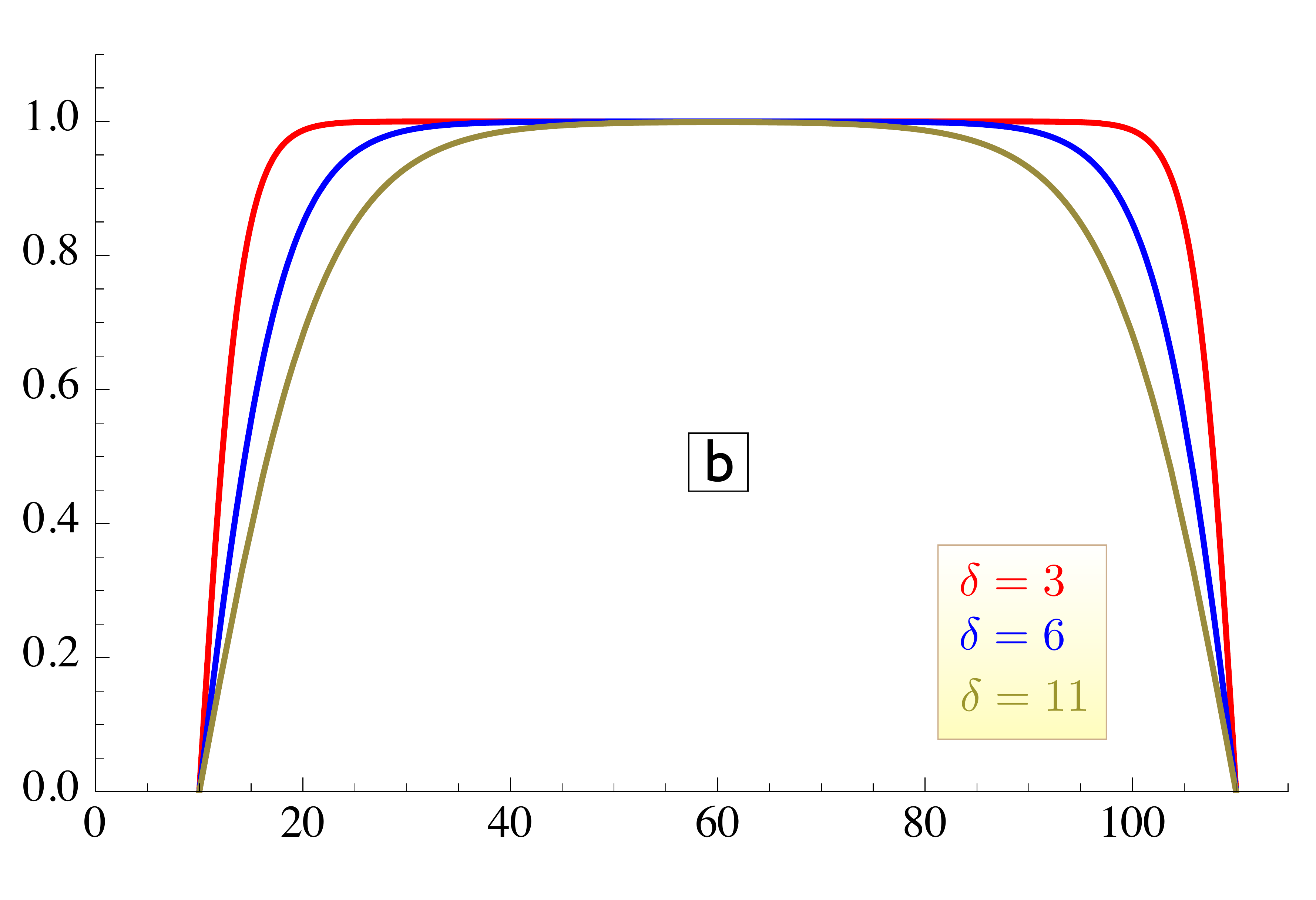}\includegraphics[width=.30\textwidth]{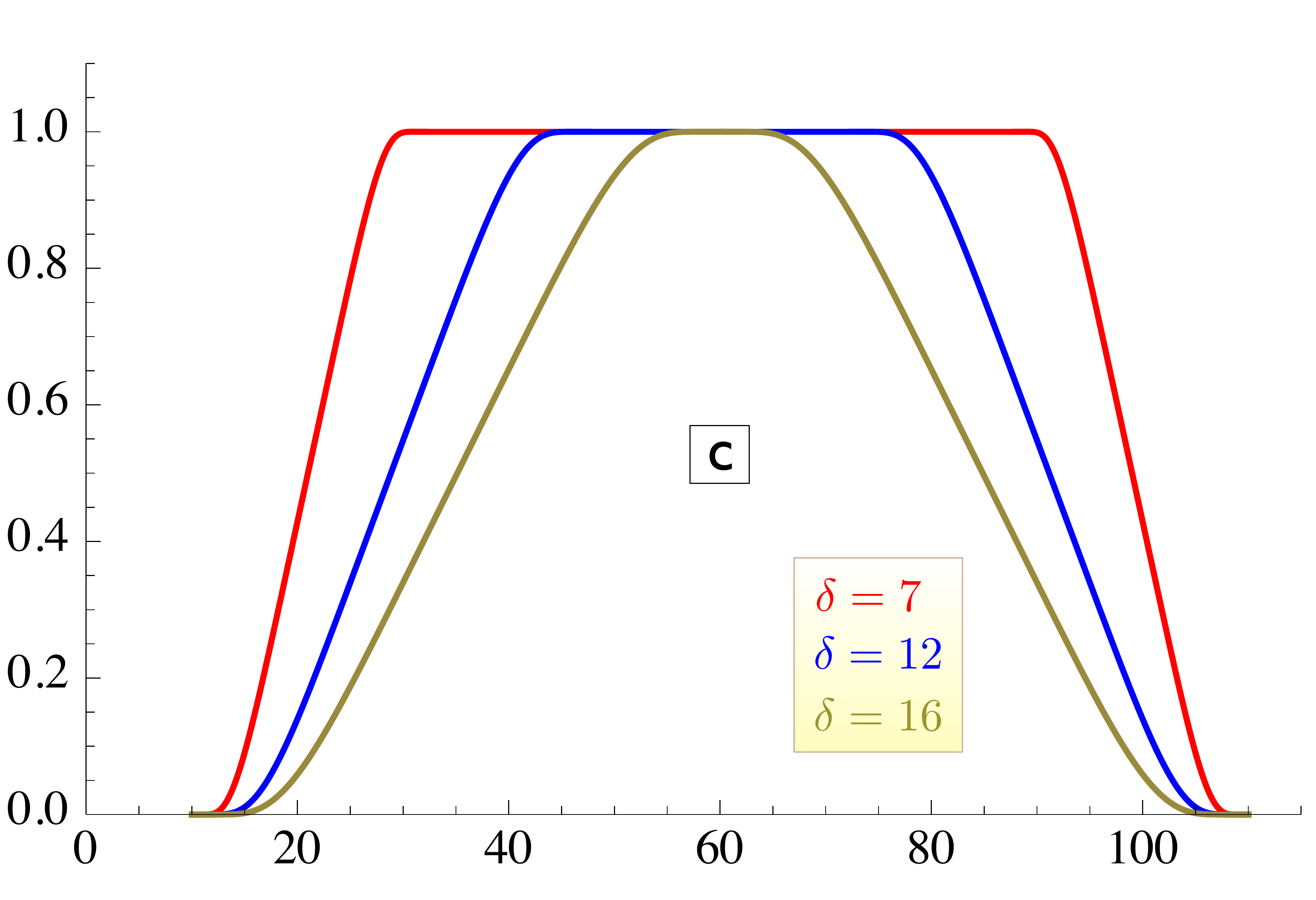}
\caption{\label{fig:switching}Compactly supported continous  switching functions defined from $t=T_0=10$ to $t=T=110$. Following function \eqref{redios1}, in {\bf [a]}, the detector is switched on at time $T_0$ then the intensity of the coupling increases linearly reaching a constant value, then ramps down to zero linearly.  Following function \eqref{redios3}, in {\bf [b]}, the detector is switched on at time $T_0$ and ramps up following an analytical switching function that quickly tends to a constant value and that decreases down to zero near the final time $T$ in the same manner. Following function \eqref{redios4}, in {\bf [c]} the detector is analytically swtiched on more smoothly  than in the case a), being the switching function compactly supported and analytical in all its domain including $t=T_0$ and $t=T$.   The scale of the ramping up and down of the switching is controlled via the characteristic time parameter $\delta$ as shown in  \eqref{redios1}, \eqref{redios3} and \eqref{redios4}. The Y-axis in these plots is expressed in units of the coupling strength~$\lambda$. }
\end{center}
\end{figure*}

 Remarkably, the difference between the long time averaged response of highly
sub-Planckian detectors in the GR and LQC scenarios \eqref{estimator} remains non-negligible even 
under these coarse-graining conditions.  As shown in Fig. \ref{fig:tomajeroma}, the variation of the response of the detector (the 
intensity of Gibbons-Hawking-type quantum fluctuations) grows exponentially with the size of the quantum of length 
$l$.
Notice that in this figure, several different switching functions have  been employed---namely,
\begin{eqnarray}
\label{redios} \chi_1(t)&=1, \qquad \\[3mm]
\label{redios1}\chi_2(t)&=\left\{\begin{array}{ll}
{\min\left[ ({t-T_0})/{\delta},1\right]}& \qquad   t<(T+T_0)/{2} \\[3mm]
{\min\left[({T-t})/{\delta},1\right]}& \qquad t\geq(T+T_0)/{2},
\end{array}\right.\\[3mm]
\label{redios3} \chi_3(t)&= \tanh \left(\textstyle{\frac{t-T_0}{\delta}}\right)-\tanh \left(\textstyle{\frac{t-T}{\delta}}\right)+\tanh \left(\textstyle{\frac{T_0-T}{\delta}}\right), \\[3mm]
\label{redios4}\chi_4(t) &= \left\{ \begin{array}{ll}
S\left[({t-T_0})/{\delta} \right] \qquad&    t<T_0 +\pi \delta\\[2mm]
1&  t\in[T_0   +\pi \delta, T - \pi \delta) \\[2mm]
S\left[ ({T-t})/{\delta}\right] &   t\ge T-\pi \delta ,
\end{array}\right.
\end{eqnarray}
where $S(x)=[1-\tanh(\cot x)]/2$ and $\delta$ controls the ramping up. These are illustrated in Fig.~\ref{fig:switching}, 

As a consequence of this exponential trend, $l$ cannot be much beyond the Planck scale or the effects would be too large nowadays. This suggests that cosmological observations could put stringent upper bounds to  the size of the quantum of volume in LQC or, equivalently, to the time scale $T_\mathrm{m}$ when the quantum dynamics corrections become  negligible.

Notice that in Fig. \ref{fig:tomajeroma} we see how the effect does not depend on the timescale of the detector's activation or the nature of  the switching function~\cite{echo}.   It is known that in $3+1$D, the sudden switching $\chi_1(t)$ this leads to UV divergent integrals  that  only depend on the switching and not on the state of the field or the background geometry~\cite{Louko:2007mu}. Nevertheless, as the authors are computing differences in probabilities of detectors with the same switching functions, their results are devoid of any such switching effect  (including UV divergences). That is the reason why   $\chi_1(t)$ also produces the same results as the continuous switchings.

Although, obviously, this is  a rather simplified toy model, it captures the essence of a key phenomenon:  Quantum field fluctuations are extremely sensitive to the physics of the early Universe. More importantly, the signatures of these fluctuations survive in the current era. 

It is important to emphasise that  the use of LQC in this derivation is rather irrelevant: The authors assert that their result is far more general, and they use LQC just as a convenient example of an early Universe dynamics model different from GR.  For any other model there has to be intermediate energy scales where a semi-classical theory is applicable, producing effective quantum corrections to the Friedmann equations. This will predict, for a short time, a spacetime dynamics which deviates from GR through quantum corrections. The main result in \cite{echo} prevails: The response of a particle detector today carries the imprint of the specific dynamics of the spacetime in the early Universe.

 It has been recently discussed \cite{robort2} that the violation of the strong Huygens effect (see \cite{czapor2007hadamard}) in curved spacetimes can give raise to the propagation of information within the timelike region of the light cone even if the field through which information propagates is a massless one. Inspired by the results on entanglement harvesting and relativistic quantum communication~\cite{Reznik2005,Robort,PastFutPRL,robort2}, current research derived from the results summarised above~\cite{echo} aims toward extending this methodology to further explore this combination of cosmology and quantum information---in particular, to study the bounds to  optimal transmission and recovery of  information propagated through cosmological catastrophes such as the Big Bang, inflation or a quantum bounce \cite{Inprepblasco}.

\section{Entanglement harvesting}
\label{sec:entharvest}

\subsection{Violating Bell inequalities in vacuum}
\label{subsec:vacuumbell}

As discussed in Sec.~\ref{sec:fieldent}, the vacuum of a quantum field is entangled with respect to local degrees of freedom. But in that section it was also stressed that the presence or absence of entanglement does not necessarily dictate observable effects of laboratory experiments.  In 2003, Reznik~\cite{Reznik2003} bridged this gap by showing that field entanglement can be extracted by local quantum systems interacting with the field in a spacelike separated way. In that proposal, two quantum systems (qubits) begin in their respective ground states and interact with the field locally for a short time. These interaction events are spacelike separated to ensure that any entanglement obtained has been swapped from the field (since entanglement cannot be increased solely by local operations and classical communication~\cite{Vidal2002}).

Having established that entanglement can be \emph{harvested} from quantum fields a natural next question to ask is, ``What can it be used for?'' Reznik and coauthors showed in 2005~\cite{Reznik2005} that harvested entanglement admits a practical use---in particular, demonstrating violation of a Bell inequality. While not particularly useful for any application, this result at least established that naturally occurring field entanglement is ``real'' entanglement in the sense of demonstrating a testable violation of local causality~\cite{Bell1964}. Subsequent work showed that this entanglement is sensitive to many parameters---sometimes highly so. These include the background spacetime and state of the field~\cite{VerSteeg2009}, the type of field~\cite{Nambu:2013gx}, and the relative separation and state of motion of the detectors~\cite{Salton:2014vj}. While each of these results suggests a potential application, none of them is immediately practical. One of the main problems is the exceptionally small amount of entanglement that can be harvested under the condition of spacelike separation.

Having the detection events spacelike separated was useful as a foil for skeptics who might wonder if the entanglement in the field is really being harvested or if the field is simply a medium for performing a nonlocal operation, which directly prepares an entangled state. If it were the latter, then the question of the physicality of a field's vacuum entanglement would remain open. This physicality already having been established by previous results with strict spacelike separation~\cite{Reznik2003,Reznik2005}, further applications of entanglement harvesting might wish to relax the condition of spacelike separation in order to boost the strength of the quantum correlations to detectable levels.

One particularly promising method of entanglement harvesting represents an evolution from a hunter-gatherer approach to an agricultural one: In so-called \emph{entanglement farming}~\cite{farming}, pairs of detectors interact sequentially with a field in a cavity, each for a short time (but not spacelike separated), harvesting entanglement and also modifying the field state as each one does so. Eventually, the field state reaches a (long-time meta-stable) fixed point from which additional detector pairs will harvest entanglement at a constant rate, each one exiting the cavity in the same state. This fixed-point state is robust to small perturbations in the initial conditions, justifying it being called `farming'. Inspired by the results from Ref.~\cite{Salton:2014vj} that show that harvested entanglement is highly sensitive to distance, this farming approach was adapted in recent work~\cite{Brown:2014uj} to serve as a ``quantum seismograph'' capable of detecting vibrations of the enclosing cavity through a change in the harvested entanglement. The rest of this section summarises recent progress in these directions.

\subsection{Entanglement harvesting with qubit detectors}
\label{subsec:harvestqubit}

To date, in all entanglement-harvesting scenarios using qubits as detectors~\cite{Reznik2003,Reznik2005,VerSteeg2009,Nambu:2013gx,Salton:2014vj}, results are obtained using an Unruh-DeWitt coupling, Eq.~\eqref{eq:UdW}, between the field and detector. The detectors have identical energy gaps~$\Omega$ and are assumed to have identical switching functions~$\chi(\tau)$ [see Eq.~\eqref{eq:UdW}], which time-dependently moderate the overall Unruh-DeWitt coupling strength~$\lambda \ll 1$ as a function of proper time~$\tau$. In the next two subsections, the switching function is taken to be a Gaussian with standard deviation~$\sigma \ll L$, where $L$~is the separation between the detectors. Therefore, the detection events are not strictly spacelike separated, although the nontrivial part of the detection events are so.

Applying the positive partial transpose criterion~\cite{Peres1996}, which is necessary and sufficient to prove separability of two qubits~\cite{Horodecki1996}, to the reduced density matrix of the detectors, one can show that the detectors are entangled if and only if~\cite{Reznik2005,VerSteeg2009,Salton:2014vj}
\begin{equation}
\left| X \right| > A\,,
\end{equation}
where $A$~is the probability of exactly one of the detectors to be excited, and $X$~is an off-diagonal term in the density matrix that corresponds to a Casimir-like effect~\cite{robort2} between the two detectors. This is sometimes referred to as an amplitude for exchange of virtual particles between them~\cite{Reznik2005}.  Up to~$O(\lambda^2)$, these are explicitly
\begin{equation}
A=\lambda^2 \int_{-\infty}^{\infty}d\tau\int_{-\infty}^{\infty}d\tau'\chi(\tau)\chi(\tau')e^{-i\Omega(\tau-\tau')}D^+(x(\tau);x(\tau'))
\end{equation}
and
\begin{equation}
\label{eq:Xeq}
	X = -\lambda^2\int_{-\infty}^{\infty}d\tau\int_{-\infty}^{\tau}d\tau'\chi(\tau)\chi(\tau')e^{i\Omega(\tau+\tau')} [D^+(x_{a}(\tau);x_{b}(\tau'))+D^+(x_{b}(\tau);x_{a}(\tau'))],
\end{equation}
where $D^+(x;x') = \langle\phi(x)\phi(x')\rangle$ is the Wightman function for the field $\phi$. In the case of the Minkowski vacuum, this is given by Eq.~(3.59) of Ref.~\cite{Birrell1982}:
\begin{equation}
D^+(x(\tau);x'(\tau'))=\frac{-1}{4\pi^{2}[(t-t'-i\epsilon)^{2}-|\mathbf{x}-\mathbf{x}'|^{2}]},
\label{eq:generalDPlus}
\end{equation}
and we set $\epsilon \to 0^+$ at the end of all calculations. (For other spacetimes and/or other field states, this has a different form~\cite{VerSteeg2009,Nambu:2013gx,Salton:2014vj}.) Note that the Wightman function as used in~$X$ contains the trajectories for both detectors, $x_a(\tau)$ and $x_b(\tau)$, while $A$ only makes use of a single detector's trajectory, applying equally well to either~$x_a(\tau)$ or~$x_b(\tau)$, due to symmetry.

The differences in the harvested entanglement result from different detector trajectories [captured in $x_{(a,b)}(\tau)$] and different field states and background spacetime [captured in $D^+(x;x')$]. In all cases, the authors assume a massless, conformally coupled scalar field in (3+1)-dimensional spacetime.

\subsection{Entangling power of an expanding universe}
\label{subsec:entpower}

Starting from the discovery by Reznik that entanglement can be harvested from a field by local detectors~\cite{Reznik2003} and used as a resource~\cite{Reznik2005}, Ver~Steeg and Menicucci in 2009 proposed to detect expansion of the universe through its entanglement-harvesting signature~\cite{VerSteeg2009}. This was interesting because it proposed to link a usable quantum information-theoretic resource (harvested entanglement) with an aspect of general relativity and cosmology (de~Sitter expansion).

The authors consider two detectors on comoving trajectories (and therefore inertial) in a de~Sitter universe with expansion rate~$\kappa$. Each detector perceives thermal radiation identical to the response that would also be obtained if the spacetime were Minkowski and the field had just been heated up to an actual thermal state in some rest frame, with both detectors at rest in that same frame. As such, a single inertial detector cannot distinguish an empty but exponentially expanding universe from a heated one that is not expanding. The authors proceed to show that when the detectors are used together to harvest entanglement, however, there is a detectable difference that can be used to distinguish the two cases.

For the expanding case, a conformally invariant setup (conformal field and conformal vacuum) was chosen for calculational simplicity, but this also has a physical significance as not producing any actual particles if the expansion were to cease later on (see Sec.~\ref{nonstaint}). In this way, it is the most conservative choice in terms of minimal particle production. More recently, Nambu~\cite{Nambu:2013gx} showed that minimal coupling instead of conformal coupling leads to a very different entanglement-harvesting profile. Since minimal coupling is used in many inflationary models~\cite{Peiris2003}, this is an important case to consider. In what follows, we just summarise Ref.~\cite{VerSteeg2009} and refer the reader to Ref.~\cite{Nambu:2013gx} for more details on the minimally coupled case.

It should be noted that the conformal vacuum also coincides with the massless limit of the adiabatic vacuum for de~Sitter spacetime~\cite{Birrell1982}. Thus, the physical setup is equivalent to the following two modifications of the Minkowski vacuum: (1) very slowly heating the universe to a temperature~$T$, and (2) very slowly ramping up the de~Sitter expansion rate (from zero) to a final value of~$\kappa$, with $T = \kappa/2\pi$. In both cases, the detectors are activated only long after this smooth transition is complete, and each detector responds as if it were bathed in thermal radiation in its own rest frame. The Gibbons-Hawking effect~\cite{Gibbons1977} is used to link the perceived temperature in the de~Sitter case with the actual temperature in the thermal case so that access to one detector alone cannot distinguish the two cases.

\begin{figure}[!tb]
\begin{centering}
\includegraphics[width=0.45\columnwidth]{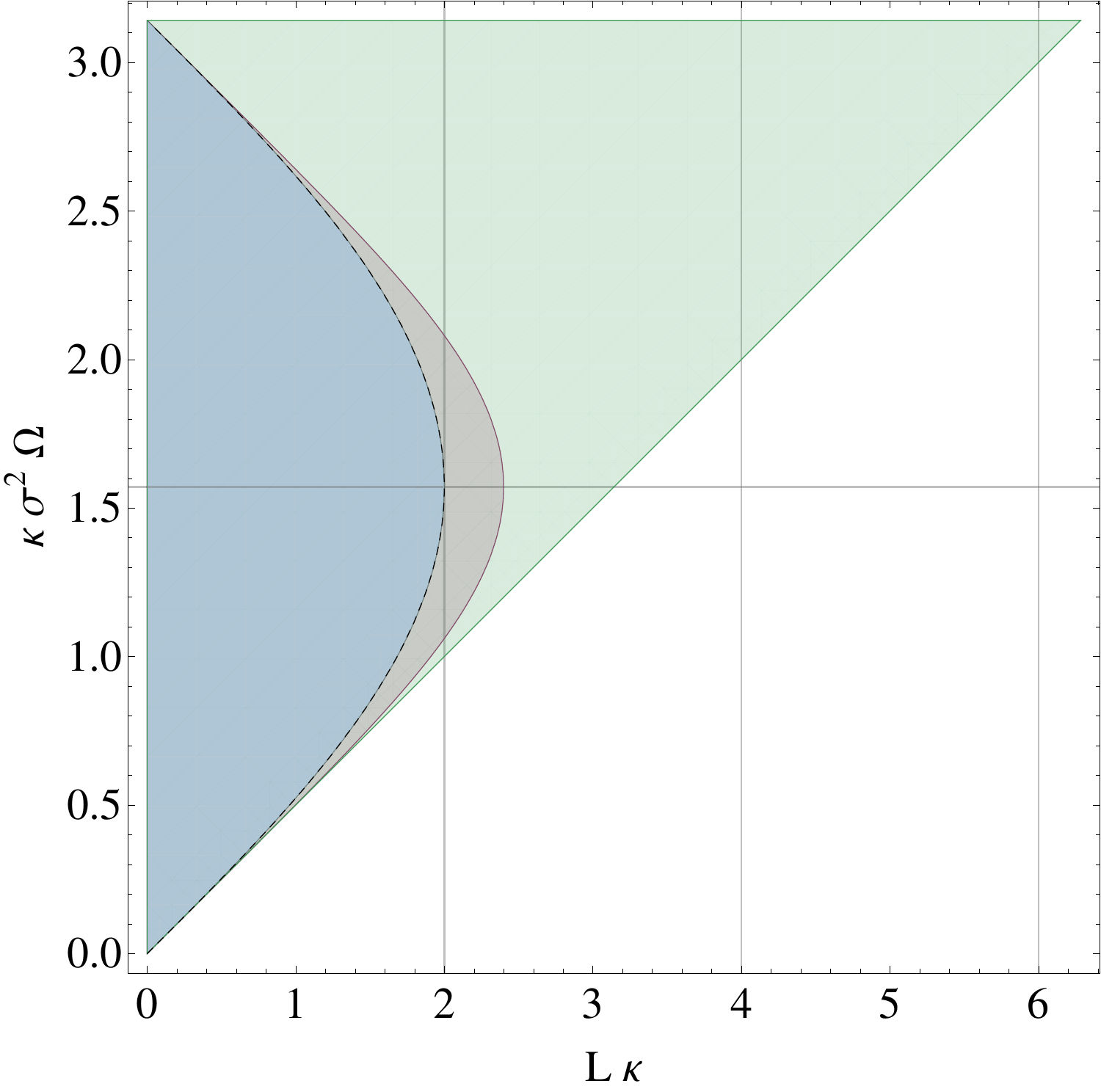}~\includegraphics[width=0.45\columnwidth]{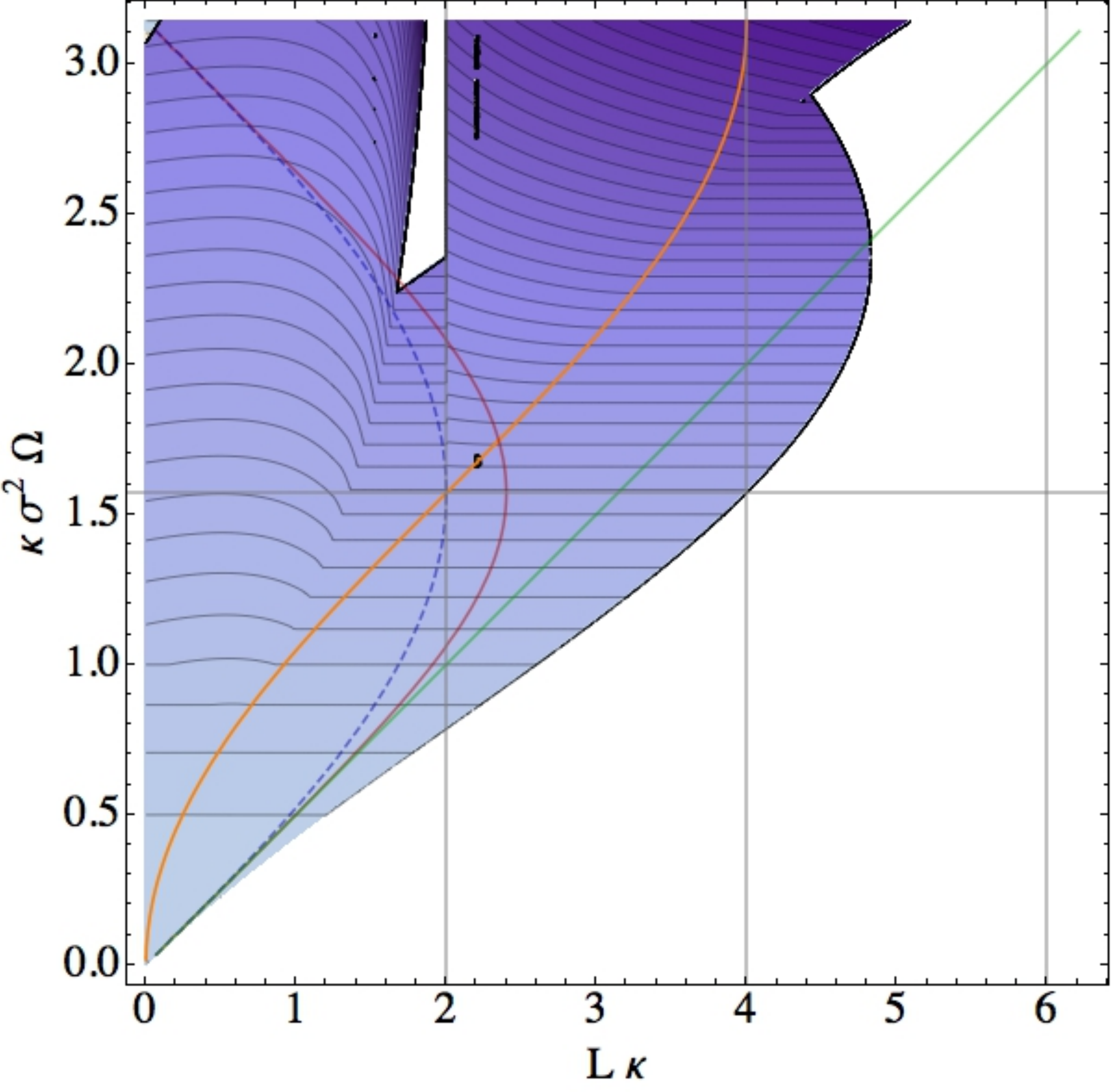}
\par\end{centering}
\caption[Negativity contours for parallel and anti-parallel acceleration]{\label{fig:negContours}[Adapted from Ref.~\cite{Salton:2014vj}] \textbf{(Left)} Allowed regions of entanglement harvesting; see text for meaning of parameters and details of physical setup. Blue:\ comoving detectors in de~Sitter conformal vacuum (Sec.~\ref{subsec:entpower}) and also parallel accelerating detectors in Minkowski vacuum (Sec.~\ref{subsec:accelharvest}). Red+blue:\ inertial detectors in Minkowski thermal bath at same perceived temperature. Green+red+blue:\ inertial detectors in Minkowski vacuum ($\kappa \to 0$). \textbf{(Right)} Entanglement profile for anti-parallel accelerated detectors. Harvesting is possible in the blue region. Blue, red, and green lines are corresponding borders of the regions in the left panel. The (mostly) horizontal contours are lines of constant log-negativity~\cite{Vidal2002} of harvested entanglement, with more entanglement toward the bottom. The four features discussed in Sec.~\ref{subsec:accelharvest} are illustrated as follows: (1)~The portion of the blue region below the green line is the region of enhancement over inertial detectors. (2)~Orange line shows the critical distance, Eq.~\eqref{eq:Lcrit}, for entanglement resonance. (3)~The curved bump on the left-hand side is due to the causal residue contribution. (4)~The triangular region in the upper half and with $L\kappa > 2$ corresponds to the noncausal residue contribution.
}
\end{figure}

Using detectors as in Sec.~\ref{subsec:harvestqubit} above, the regions in parameter space for which entanglement harvesting is possible by comoving detectors in the three cases of expanding, heated, and the Minkowski vacuum ($T\to 0$ limit of the other two) can be found either numerically~\cite{VerSteeg2009} or analytically using a stationary phase approximation~\cite{Nambu:2013gx,Salton:2014vj}. For the expanding case, entanglement can be harvested if and only if
\begin{equation}
\label{eq:dSentcond}
    \frac {L\kappa} {2} < \sin (\kappa\sigma^2\Omega)\,.
\end{equation}
In the case of a thermal Minkowski universe, the condition for entanglement harvesting is
\begin{equation}
    \frac {L\kappa} {2} \tanh\left(\frac{L\kappa}{2}\right)<\sin^2(\kappa\sigma^2\Omega)\,.
\end{equation}
The boundary for the Minkowski vacuum can be found by taking the $\kappa \to 0$ limit of either of these:
\begin{equation}
    \frac L 2 < \sigma^2\Omega\,.
\end{equation}
These regions are shown in the left plot of Fig.~\ref{fig:negContours}. Notice that there exists a region of parameter space in which a heated universe entangles the detectors while an expanding universe at the same perceived temperature does not. As such, the two types of universes can be distinguished by their ``entangling power'' with respect to local inertial detectors.

In both the thermal Minkowski and de~Sitter vacuum cases, the region of entanglement is a proper subset of that of the Minkowski vacuum case. In the thermal case, this is expected because the thermal state has higher entropy, but this is somewhat surprising for the expanding case since the state is still a type of vacuum. 

For the expanding case, we consider the behavior of the entire system in conformal time rather than in the detectors' proper time. Since the field is conformally invariant, the modes behave like ordinary Minkowski modes with respect to conformal time~\cite{Birrell1982}. To be resonant with these modes, a detector would need to have a fixed resonance frequency \emph{in conformal time}, which has the same dynamics as those of a detector with a changing frequency in proper time. Were the detectors tuned in this nonstandard way, one could recover all the entanglement of the Minkowski vacuum (simply by being resonant with the same modes in conformal time rather than ordinary time). Ordinary detectors, however, have a fixed frequency in proper time, which corresponds to a time-varying frequency with respect to conformal time. Therefore, the detectors are time-dependently sweeping in and out of resonance with individual Minkowski modes due to the expansion. This effect gives rise to greater noise (i.e.,~larger~$A$) in each detector due to the counter-rotating terms in the Unruh-DeWitt coupling, Eq.~\eqref{eq:UdW}, and this acts to reduce the harvested entanglement.

Notice that the difference between the expanding and thermal cases only really shows up when $L\kappa > 1$, which is precisely when the detectors are separated~($L$) by more than a cosmic horizon distance~($\kappa^{-1}$). Thus, the detectors cannot communicate with each other to verify their entanglement, nor can they use the entanglement for quantum teleportation or long-distance quantum communication with each other since teleportation still requires classical communication of local measurement results~\cite{Bennett1993}.

Imagine, however, that the detectors are mounted on satellites that can communicate with a ``home planet'' half-way between them. Assuming many repetitions of (attempted) entanglement harvesting, followed by a local measurement protocol to verify separability or entanglement, which is always possible~\cite{Masanes2006,Horodecki1997}, classical data can be sent to this home planet for verification. Furthermore, given an initial source of entanglement between the home planet and each satellite separately, any entanglement harvested from the quantum field could be swapped via teleportation back to the home planet~\cite{Bennett1993, Zukowski1993}. This method of remote entanglement harvesting is impractical since the harvested entanglement is way too small~\cite{VerSteeg2009}, but the thought experiment shows that in principle the existence of this entanglement is verifiable since it can be used as a quantum information processing resource back on the home planet.

\subsection{Acceleration-assisted entanglement harvesting and rangefinding}
\label{subsec:accelharvest}

There is an elegant mathematical connection between Gibbons-Hawking radiation~\cite{Gibbons1977} and the Fulling-Davies-Unruh effect~\cite{Fulling1973,Davies1975,Unruh1976}, so a natural question to ask is what happens if we try to do entanglement harvesting using accelerated detectors through the Minkowski vacuum. Salton, Mann, and Menicucci~\cite{Salton:2014vj} tackle this question, with rather surprising results.

\begin{figure}[!tb]
\begin{centering}
\includegraphics[width=.30\columnwidth]{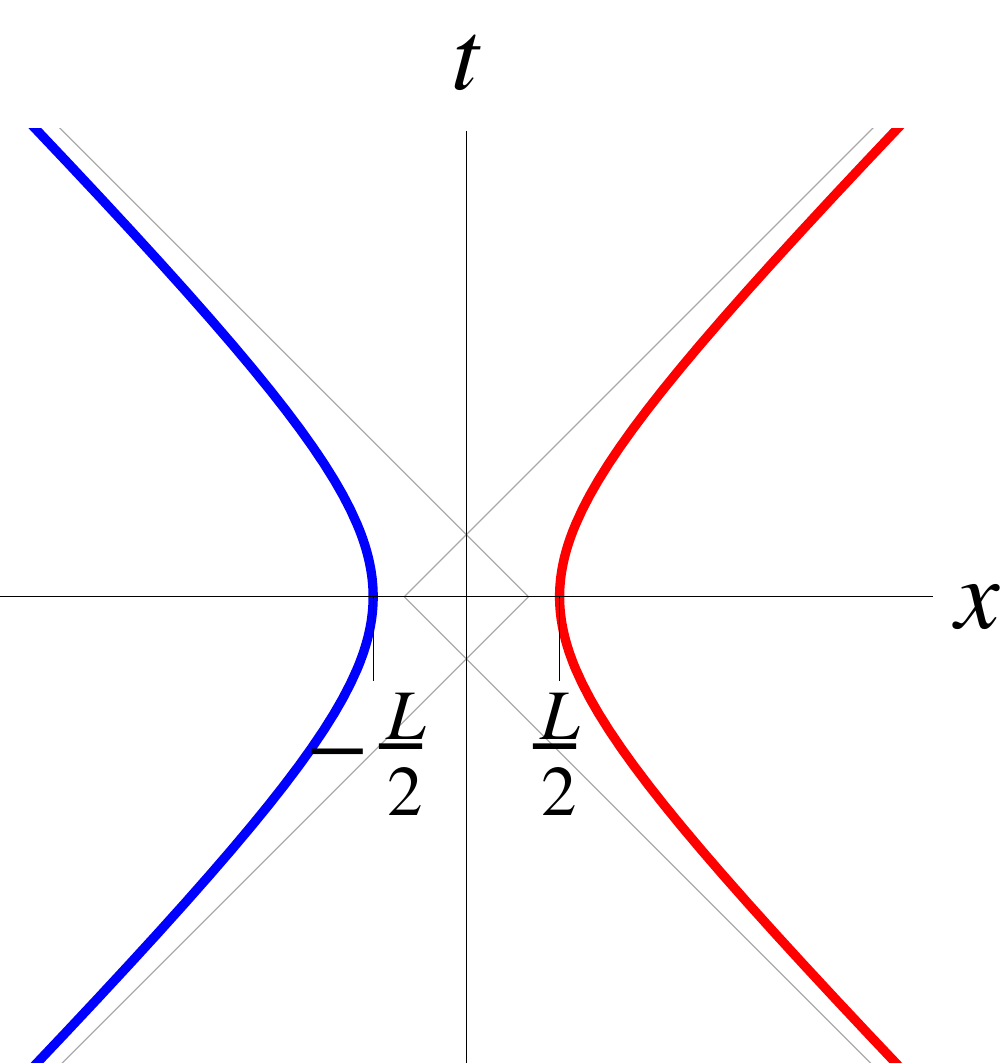}\qquad
\includegraphics[width=.30\columnwidth]{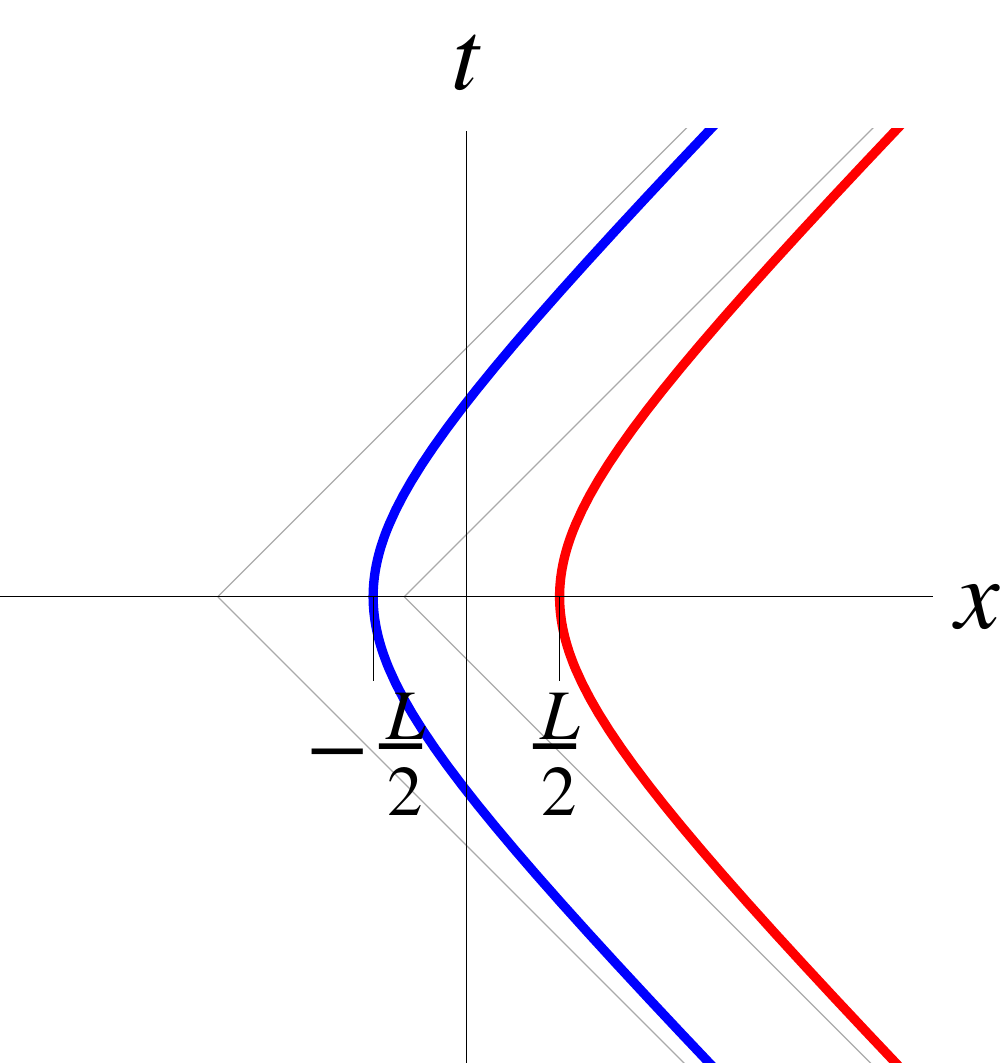}
\par\end{centering}
\caption[Worldlines of two uniformly accelerated detectors]{\label{fig:Trajectories}[Adapted from Ref.~\cite{Salton:2014vj}] Worldlines of two detectors undergoing anti-parallel (left panel) or parallel (right panel) uniform acceleration of equal magnitude. The grey lines indicate the Rindler wedges associated with each trajectory.}
\end{figure}

The premise of that work~\cite{Salton:2014vj} is simply to replace inertial trajectories with uniformly accelerated ones. The authors choose two cases: parallel and anti-parallel acceleration (of the same magnitude for both detectors):
\begin{eqnarray}
x_a = \phantom\pm\frac{1}{\kappa}\Bigl[\cosh(\kappa\tau) - 1\Bigr] +\frac{L}{2}%
\,, &\qquad& 
t_a = \frac{1}{\kappa}\sinh(\kappa\tau)\,, \\
x_b = \pm\frac{1}{\kappa} \Bigl[ \cosh(\kappa\tau) - 1 \Bigr] -\frac{L}{2}%
\,, &\quad&
t_b = \frac{1}{\kappa}\sinh(\kappa\tau) \,,
\end{eqnarray}
where the minus/plus signs refer to anti-parallel/parallel trajectories, respectively illustrated in left/right panels of Fig.~\ref{fig:Trajectories}. Importantly, and in contrast to previous work~\cite{Massar2006,Birrell1982,Unruh1976}, the Rindler wedges (for each observer) in the anti-parallel case do not necessarily share a common apex. Instead, they could be closer (as shown) or further apart. The distance of closest approach of the detectors, as measured by an inertial observer at fixed~$x$, is $L$.

The switching functions and all other parameters and methods are the same as in Sec.~\ref{subsec:harvestqubit}. To calculate $A$ and $X$, the authors use complex analysis, shifting the integration contours in the imaginary direction and using the stationary phase approximation. This reveals that in the case of parallel acceleration, entanglement can be harvested if and only if
\begin{equation}
    \frac {L\kappa} {2} < \sin (\kappa\sigma^2\Omega)\,.
\end{equation}
Curiously, this is the exact same condition as that of de~Sitter expansion, Eq.~\eqref{eq:dSentcond}, and thus the entangling region is already shown in the blue region of the left panel of Fig.~\ref{fig:negContours}.

The anti-parallel case turns out to have a much richer structure. While $A$ is the same as in the parallel case, $X$ turns out to be much more complicated. It includes additional contributions from poles that are crossed when shifting the contour (as described in the parallel case). Furthermore, there is one prominent feature (entanglement resonance, described below) that gets missed when employing the stationary phase approximation; numerical integration is required in the vicinity of a narrow corridor in parameter space. These features together mean that no simple inequality suffices to describe the entanglement profile in parameter space of the anti-parallel case. Instead, the results are simply shown in the right panel of Fig.~\ref{fig:negContours}, which also contains the outline of the boundaries of the regions shown in the left panel for reference. There are four main features of interest in this case, which we describe presently.

\subsubsection*{(1)~Enhancement of harvesting over inertial detectors}

The first region of interest is the part of the entangling region that extends below the limit for inertial detectors (green line). This feature has a curious physical interpretation. Two inertial detectors in Minkowski spacetime separated by $L \gtrsim 2\sigma^2\Omega$---i.e., just below the green diagonal line in the right panel of Fig.~\ref{fig:negContours}---cannot harvest entanglement. But if they were to fly in toward each other from spatial infinity, all the while accelerating in the opposite direction and eventually retreating back to infinity, there exist resonance frequencies such that they could harvest entanglement even if their distance of closest approach~$L$ is larger than $2\sigma^2\Omega$. Notice that we do not have to physically bring the detectors closer together than the original~$L$ in order to entangle them; the anti-parallel acceleration is sufficient to nudge them into an entangling regime in parameter space.

Note that in the limit of no acceleration, $\kappa \to 0$, the minimum values of $L$ and $\Omega$ required to see this enhancement would both become infinite, which is impossible. Therefore, the inertial condition for entanglement ($L < 2\sigma^2\Omega$) becomes the appropriate one for all practical purposes in this case, which is a good consistency check.

\subsubsection*{(2)~Entanglement resonance}

The orange line in the right panel of Fig.~\ref{fig:negContours} represents a \emph{critical distance}
\begin{equation}
\label{eq:Lcrit}
	L_{\text{crit}}=\frac{2}{\kappa}\left(1-\cos(\kappa\sigma^2\Omega)\right)
\end{equation}
associated with resonance frequency~$\Omega$ such that the coherence term~$X \to \infty$. This indicates a breakdown in perturbation theory (since the density matrix is no longer positive), which the authors interpret as being due to a resonance effect---i.e., the field inducing a strong dynamical effect on the detectors when they have those particular parameters, possibly allowing for a large amount of entanglement to be harvested. With the analysis limited to second order perturbation theory, the authors leave verifying this hypothesis to future work.

The existence of a resonance in the anti-parallel case is not in itself surprising, considering the Unruh effect can be considered to be a result of two-mode squeezing between field modes isolated to left and right Rindler wedges~\cite{Unruh1976}, as shown in Sec.~\ref{Rindbogo}. One would expect that appropriately arranged detectors might be resonant with the two-mode-squeezed modes and thereby harvest a large amount of this Rindler entanglement. What the usual formalism suggests~\cite{Birrell1982,Unruh1976}, however, is a natural relationship between $L$ and $\kappa$ so that the asymptotes associated with the two trajectories all meet at a single point and form a single set of Rindler wedges. This occurs when $L\kappa = 2$, but the authors of Ref.~\cite{Salton:2014vj} do not restrict to this case. In fact, there is a resonance effect at $L\kappa = 2$ only if $\kappa \sigma^2 \Omega = \frac \pi 2$. In addition, however, there is also a resonance effect for any $L\kappa < 4$ if one chooses an appropriate~$\Omega$. That such a resonance occurs for a wide range of separations is surprising and demonstrates that the interplay between the detectors and the field entanglement is nontrivial.

\subsubsection*{(3)~Causal residue contribution}

As mentioned above, evaluating $X$ involved residue contributions from poles crossed when shifting the integration contour. There are two distinct types of residue contributions. The first type is a contribution that is largest for small~$L\kappa$ and vanishes when $L\kappa > 2$. Due to this behavior, the authors suggest that the residue contribution in this region may arise from causal dynamics. This contribution can be seen as the curved ``bulge'' in entanglement on the left-hand side of the right panel of Fig.~\ref{fig:negContours}.

When $L\kappa < 2$, the infinite tails of the Gaussian switching function guarantee that the detection events have a causal overlap despite the fact that the interaction strength of at least one of the detection events is extremely weak in that overlap region. Although the interaction is very weak, the long interaction time with the underlying field could give rise to a nontrivial contribution to $X$ that is comparable to (or even greater than) the residue-free portion.

Specifically, the authors suggest that a detector's interaction with the field for a long time in the infinite past may produce a large effect concentrated near the lightlike past asymptote of that detector's trajectory that is felt by the other detector (as a contribution to the coherence term~$X$) as the second detector crosses the future extension of that asymptote. Additional effects generated by exchanging the detectors and by considering the time-reversed process contribute similarly. 

The net result of these processes is a nontrivial contribution to the coherence term~$X$, which indicates that the detectors ``feel each other's presence'' due to the long tails of the Gaussian and overlapping wedges.

\subsubsection*{(4)~Noncausal residue contribution}

Even as the causal residue contribution discussed above vanishes as $L\kappa > 2$, a second residue contribution takes over for those greater distances---but only when $\kappa\sigma^2 \Omega > \frac \pi 2$. This means that there exists a nontrivial residue contribution for purely spacelike separated detectors ($L\kappa>2$) but only for certain choices of the other parameters.

To interpret this contribution, the authors note that the effect is strongest near $L\kappa \gtrsim 2$ and argue that this is evidence that the detectors may be feeling the effects of the entangled Rindler modes~\cite{Unruh1976,Birrell1982}. Notice that if $L\kappa \gtrsim 2$, the detectors will be nearly resonant with the Rindler modes that are two-mode squeezed~\cite{Unruh1976,Birrell1982}. The trajectories align perfectly as $L\kappa \to 2^+$. The fact that this contribution, mathematically, cuts off sharply as soon as $L\kappa < 2$ also suggests that the infinite tails may play a role in this effect.

Because the methods used to calculate $X$ in Ref.~\cite{Salton:2014vj} rely on the switching function being analytic, further work will be required to determine which of these effects survive if the window functions are made strictly compact in proper time.

\subsubsection*{Rangefinding}

The authors of Ref.~\cite{Salton:2014vj} suggest several means by which entanglement harvesting could---as a thought experiment---be used to measure the distance of closest approach between two detectors undergoing antiparallel acceleration. The basic idea is to tune the detector parameters such that they are in a regime where the harvested entanglement changes rapidly with small changes in~$L$. There are several possibilities for this. The first is the resonance (orange line in Fig.~\ref{fig:negContours}), for which the resonance effect is in an extremely narrow region in $L$ around the critical distance, Eq.~\eqref{eq:Lcrit}. In fact, the sign of~$X$ changes in this region, which can be detected as a flip from correlation to anticorrelation in particular joint Pauli measurements. Another proposal is to use the steep gradient in the entanglement in the causal residue portion to detect distance. Finally, the authors propose detecting the sudden onset of entanglement in the noncausal residue portion if the detectors satisfy $L\kappa \simeq 2$.

All of these approaches would require many trials where the distance would be varied slightly in each run. While these are not practical approaches due to the tiny amounts of entanglement and long distances involved, they inspired the results described the next section, which employ entanglement farming~\cite{farming} to detect vibrations in a cavity enclosing the quantum field being farmed~\cite{Brown:2014uj}.

\subsection{Entanglement farming and quantum seismology}
\label{subsec:seismo}

Making use of to the non-perturbative tools developed in Ref.~\cite{Brown}, the authors of Ref.~\cite{farming} showed that it is possible to build sustainable entanglement sources taking advantage of the build-up of relativistic effects---even in non-relativistic settings---in a reliable and experimentally accessible way. In particular, they showed that in certain generic circumstances the state of light of an optical cavity traversed by (non-relativistic) beams of atoms is naturally driven towards a non-thermal metastable state. This state can be such that successive pairs of unentangled particles sent through the cavity will reliably emerge significantly entangled, thus providing a renewable source of quantum entanglement~\cite{farming}. This entangling fixed-point state of the cavity is reached for the most part independently of the initial state in which the cavity was prepared.

This suggests that entanglement `farming'---i.e., driving the field state to a fixed point through repeated interactions with local systems---can be implemented in a broad class of quantum systems. Reliable and repeatable production of entanglement should be also possible in other experimental settings. Namely, instead of successively temporarily coupling pairs of particles to a cavity field, one may successively temporarily couple pairs or triplets, etc., of qudits to a suitable reservoir system. The qudits and the reservoir system could have any arbitrary physical realization, even outside quantum optics if the settings described are studied non-perturbatively. 

The intuition behind rangefinding from accelerated detectors~\cite{Salton:2014vj} combined with the more practical techniques of entanglement farming proposed in Ref.~\cite{farming} leads to a proposal for quantum seismology~\cite{Brown:2014uj}. In particular, it is possible to modify the entanglement farming techniques to build ultra-high-precision vibration sensors by finding suitable points at which the fixed point of the entanglement farming protocol can be made highly sensitive to perbations such as the vibration of the optical cavity walls or the passage of a gravitational wave. This makes the setup useful for ultra-precise detection of vibrations. This quantum entanglement-farming based metrology setting has been called a \emph{quantum seismograph} by the group of researchers developing it~\cite{Brown:2014uj}.

Given that it has been shown that the phenomenon of entanglement harvesting is sensitive to the spacetime background~\cite{VerSteeg2009,Nambu:2013gx} and detector motion~\cite{Salton:2014vj}, it should be interesting to study these `quantum seismograph' settings in terms of their sensitivity and noise tolerance in order to detect gravitational waves (slight perturbations of the spacetime metric)~\cite{Brown:2014uj}.%

\section{Conclusion}
\label{sec:conc}

All of these results point to an exciting interplay between quantum entanglement and cosmology, with many more insights yet to come. The fact that we can study entanglement in the fields themselves (through Bogoliubov transformations) and also with detector-based methods allows one to obtain physical insights using complementary perspectives. Recently developed nonperturbative detector models bring a new set of tools to the table and allow us to study these effects even in cases of strong coupling or other interesting non-perturbative regimes, such as the problem of equilibration-thermalization in quantum field theory.

While directly testing most predictions of curved-spacetime quantum field theory, including nearly all of the effects discussed in this work, may arguably be prohibitive, analogue models~\cite{Unruh1981,Barcelo2005} provide a means to test many of these effects in a laboratory setting~\cite{MartinMartinez2012}. More clever proposals for direct testing may yet lead to experimental discoveries on par with recent discoveries related to cosmic inflation~\cite{Collaboration:2014wq}. In either case, we hope the reader will find that the results presented here provide both intuition and concrete results into the fascinating connection between cosmology, quantum field theory, and relativistic quantum information.

\subsection*{Acknowledgments}

We thank Ana Blasco, Eric Brown, William Donnelly, Luis~J. Garay, Achim Kempf, Robert~B. Mann and Mercedes Mart\'in-Benito for inspiring discussions. This work was supported in part by the Natural Sciences and Engineering Research Council of Canada. E.M.-M.\ acknowledges the support of the NSERC Banting Postdoctoral Fellowship programme. N.C.M.\ was supported by the Australian Research Council under grant No.~DE120102204.

\vspace{1em}
\hrule
\vspace{1em}

\bibliographystyle{bibstyleNCM_papers.bst}
\bibliography{Expandingrefs,allrefs}

\end{document}